\documentclass[reprint,superscriptaddress,amssymb,aps, pra, twocolumn,graphicx]{revtex4-1}
\usepackage{lipsum}
\usepackage{subfigure}
\usepackage{amsfonts}
\usepackage{amsmath}
\usepackage{txfonts}
\usepackage{amssymb}
\usepackage{amsbsy} 
\usepackage{epsfig}
\usepackage{graphicx}
\usepackage{epstopdf}
\usepackage{geometry}
\usepackage{hyperref}
\usepackage{setspace}

\begin{document}

\title{Optomechanically induced transparency and directional amplification in a non-Hermitian optomechanical lattice }%

\author{Pengyu Wen}
\affiliation{Department of Physics, State Key Laboratory of Low-Dimensional Quantum Physics, Tsinghua University, Beijing 100084, China}

\author{Min Wang}
\email{wangmin@baqis.ac.cn}
\affiliation{Beijing Academy of Quantum Information Sciences, Beijing 100193, China}

\author{Gui-Lu Long}
\email{gllong@tsinghua.edu.cn}
\affiliation{Department of Physics, State Key Laboratory of Low-Dimensional Quantum Physics, Tsinghua University, Beijing 100084, China}
\affiliation{Beijing Academy of Quantum Information Sciences, Beijing 100193, China}
\affiliation{Frontier Science Center for Quantum Information, Beijing 100084, China}
\affiliation{Beijing National Research Center for Information Science and Technology, Beijing 100084, China}
\affiliation{School of Information, Tsinghua University, Beijing 100084, China}

\date{\today}%

\begin{abstract}

    Cavity optomechanics is important in both quantum information processing and bascic physics research.  In this paper, we propose an  optomechanical lattice which manifests non-Hermitian physics . We first use the non-Bloch band theory to investigate the energy spectrum and transmission property of an optomechanical lattice.  The generalized Brillouin zone  of the system is calculated with the help of the resultant. And the periodical boundary condition (PBC) and open boundary condition  energy spectrum are given, subsequently. By introducing probe laser on different sites we observed the directional amplification of the system. The direction of the amplification is analyzed combined  with the non-Hermitian skin effect. The frequency that supports the amplification is analyzed by considering the PBC energy spectrum.  By introducing probe laser on one site we investigate the onsite transmission property. Optomechanically induced transparency (OMIT) can be achieved in our system. By varing the parameters and size of the system, the OMIT peak can be effectively modulated or even turned into optomechanically induced amplification .  Our system shows its potential as the function of a single-way signal filter. And our model can be extended to other non-Hermitian Bosonic model which may possess topological features and bipolar non-Hermitian skin effect.    
    \end{abstract}
\maketitle

\section{Introduction}
  Hermiticity of the Hamiltonian is the basic assumption in quantum mechanics. Based on this assumption the eigenvalue of the system is real and the particle number is conserved. However, there will inevitably be particle and energy exchange between the real physics system we are intereted and the atmosphere. The non-Hermiticity is ubiquitous in nature, including the gain and loss of optical system, the friction of the mechanical system, the finite lifetime of the quasiparticle  in condensed matter physics, the dissipation of open quantum system, the measurement backaction in quantum measurement, etc\cite{ashida2020non}. In theory, the way to investigate the non-Hermitian system is to consider it as an open quantum system whose behavior is described by the Lindblad quantum master equation \cite{scully1999quantum}.  With the development of the fabrication and experimental skills, there are also plenty of experiments showing novel non-Hermitian physics, including skin effect \cite{xiao2020non,zhang2021observation,zou2021observation}, parity-time symmtry\cite{bender1998real,xia2021nonlinear,feng2017non} and topology\cite{xia2021nonlinear,xiao2017observation}.

  In a Hermitian system with translation invariance, the Bloch theorem shows its power since it describes many physical properties of the system, such as the topological invariant , the existence of  edge states and the bulk-boundary correspondence (BBC) \cite{qi2011topological,hasan2010colloquium,asboth2016short,shen2012topological} . However, the tradition BBC in Hermitian system has failed in a non-Hermitian system due to a higher sensitivity to the  boundary effect\cite{okuma2020topological,yao2018edge,yu2021generalized} . Specifically, in a Hermitian system in open boundary condition (OBC) the eigenstates are Bloch waves while in a non-Hermitian system in OBC the eigenstates are localized at the boundaries of the system with an exponential  decay into the bulk, namely the non-Hermitian skin effect  (NHSE) \cite{longhi2019probing,okuma2020topological,li2020critical}. So the non-Bloch band theory to describe the topological feature of a non-Hermitian system was established \cite{yokomizo2019non,kawabata2020non,yao2018edge,yao2018non,yang2020non,zhang2020correspondence,longhi2020non}. In this theory, the Brillouin zone (BZ) in Bloch theorem was replaced by the generalized Brillouin zone (GBZ). Accordingly, the topological invariant, e.g. the winding number was redefined as changing the integration zone from the BZ to the GBZ. Another important example of this generalization is the Green function formulas given by the non-Bloch band theory\cite{xue2021simple}.

  Optical system is  naturally a non-Hermitian system since its onsite loss and gain and nonreciprocal coupling between different optical modes \cite{feng2011nonreciprocal,bi2011chip}. Since the Maxwell equation can be written in the form of the Schrödinger equation, optical systems with translation invariance can also form the energy band, breeding the reserach of non-Hermtian effect in topological photonics\cite{ozawa2019topological,lu2014topological,smirnova2020nonlinear,khanikaev2017two,kim2020recent}.  Inspired by this, we decided to investigate a  non-Hermitian optomechanical lattice.

  Light will interact with the cavity that confines it via radiation pressure. This interaction is the well-known optomechanical interaction and has attracted wide attention among researchers over the past decade\cite{aspelmeyer2014cavity,marquardt2009optomechanics,kippenberg2008cavity,metcalfe2014applications,kippenberg2007cavity,dong2012optomechanical}. Inspired by the interaction between the optical field and the atoms, people have also been finding novel phenomena in optomechanical system. Electromagnetically induced transparency (EIT) in atom systems have found its analogy in optomechanical system, i.e. optomechanically induced transparency (OMIT)\cite{marangos1998electromagnetically,weis2010optomechanically,xiong2018fundamentals,kronwald2013optomechanically}. A traditional OMIT system consists of one optical mode and one mechanical mode. A peak in the transmission spectrum of a weak probe laser can be observed when the optical mode are coupled by a strong control field. In recent years, OMIT has been shown  in various experimental setups \cite{weis2010optomechanically,dong2013transient} . Many theoretical proposals based on OMIT in whispering gallery modes system have also  been discussed\cite{lu2018optomechanically,lai2020tunable,lu2017optomechanically,lei2015three,qin2020manipulation,mao2022tunable,peng2014parity,xie2021phase,jiang2015chip,mao2020enhanced,chen2014photon}.   OMIT is also suggested for applications such as the manipulation of light propagation\cite{jing2015optomechanically}, precision measurement\cite{zhang2012precision} and ground state cooling of mechanical motion\cite{guo2014electromagnetically,ojanen2014ground,liu2015optomechanically}.

  Realizing the directional amplification is an important issue in quantum information processing since it allows signals to propagate in a single way and strong enough to cover the noise \cite{xue2021simple,wanjura2020topological,abdo2013directional,mcdonald2018phase} . Devices such as circulators and isolators have played important roles in optical and microwave systems\cite{turner1981fiber}. In our work we utilize these elementary devices to form a long array which supports amplification for some frequency. Note that not all frequency can be amplified so that our system can also be considered as a filter.

  In this paper, we investigate the energy spectrum and the transmission property of a non-Hermitian optomechanical lattice. In section \ref{section two} we introduce our model. The non-Hermitian property of the system is introduced by not only the onsite decay, but also the nonreciprocal coupling of the adjacent optical modes. In section \ref{section two} and \ref{section three}, the non-Bloch band theory was utilized to calculate both the energy spectrum and the Green function of the system to analyze the transmission the property of the system. In section  \ref{section four} we discuss the relationship between the energy spectrum and the transmission property of the system.

\section{Model and Method \label{section two}}
\subsection{Non-Hermitian optomechanical lattice Hamiltonian}
  We begin with a finite 1D non-Hermitian optomechanical lattice shown in figure.\ref{model} below. The lattice consists of N sublattices and each sublattice is 
  composed of one optical mode $\hat{a}_{n}$ and one mechanical mode $\hat{b}_{n}$. The adjacent optical mode and mechanical mode are coupled to each other via
  standard optomechanical interaction with coupling constant $g$. At the same time the optical mode in adjacent sublattices are coupled to each other in a nonreciprocal pattern,  i.e. the left optical mode hops to the right one with strength $t-\frac{\gamma}{2}$ while the right optical mode hops to the left one with strength $t+\frac{\gamma}{2}$ . This can be easily done in experiments because theoretical proposal for nonreciprocal transport in optical system has been a widely discussed  topic in recent years\cite{turner1981fiber}. As a simplest method we can use the Faraday magneto-optical effect to achieve this. We note that this nonreciprocity will introduce dissipation to the system and is contributed to the non-Hermitian behavior of the system. Each optical mode is drived by a control laser with frequency $\omega_{L}$ and  probe channels can added to the system, e.g. in figure.\ref{model} one probe waveguide is attached to the $n$th optical mode. The optical mode and the mechanical mode has decay rate $\kappa$ and $\Gamma$, respectively. Note that the optical mode dissipation are due to both the nonreciprocal coupling inside the lattice and the coupling to the outer environment, i.e. the driving field and the vacuum bath. 
  \begin{figure}
    \includegraphics[width=9.0cm, height=4.3cm]{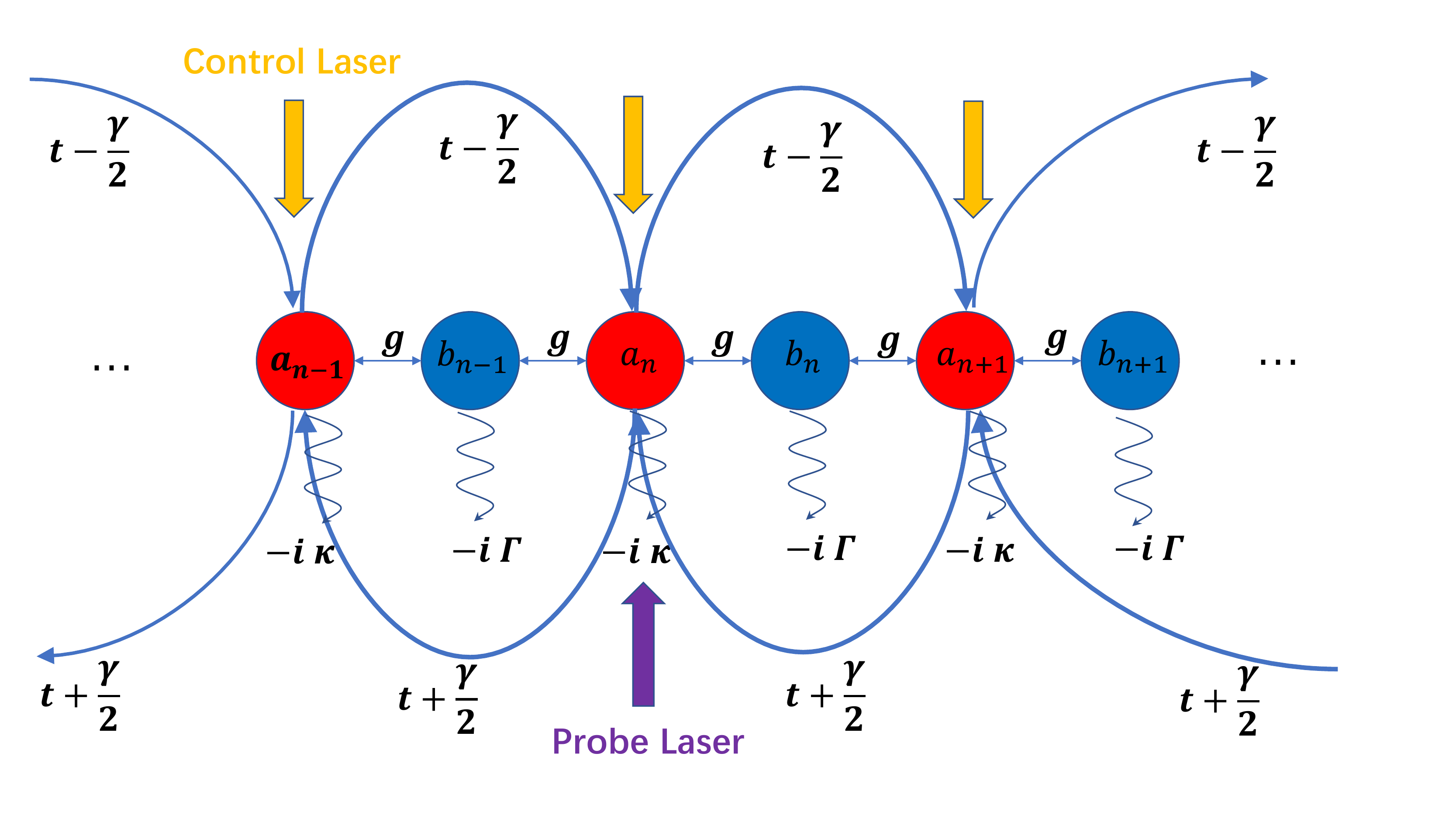}
    \caption{A finite 1D non-Hermitian optomechanical lattice. Each sublattice consists of one optical mode (red circle) and one mechanical mode (blue circle).
    The adjacent optical mode and mechanical mode are coupled to each other through standard optomechanical interaction. The adjacent optical modes are coupled to each other in a nonreciprocal way. Each optical mode is drived by a control laser and the probe waveguide is added to the site which we are intereted in.}
    \label{model}
    \end{figure}

    Our system can be described by the Hamiltonian below,
    \begin{align}
        &\hat{H}=\hat{H}_{\rm onsite}+\hat{H}_{\rm om}+\hat{H}_{\rm hop} \label{YY}\\
        &\hat{H}_{\rm onsite}=\sum_{n=1}^{n=N} (\omega_{c}-i \kappa)\hat{a}^{\dagger}_{n}\hat{a}_{n}+(\omega_{m}-i \Gamma)\hat{b}^{\dagger}_{n}\hat{b}_{n} \tag{\ref{YY}{a}} \label{YYa}\\
        &\hat{H}_{\rm om}=\sum_{n=1}^{n=N} g \hat{a}^{\dagger}_{n}\hat{a}_{n}(\hat{b}^{\dagger}_{n-1}+\hat{b}_{n-1}+\hat{b}^{\dagger}_{n}+\hat{b}_{n}) \tag{\ref{YY}{b}} \label{YYb}\\
        &\hat{H}_{\rm hop}=\sum_{n=1}^{n=N} (t-\frac{\gamma}{2})\hat{a}^{\dagger}_{n+1}\hat{a}_{n}+(t+\frac{\gamma}{2})\hat{a}^{\dagger}_{n}\hat{a}_{n+1} \tag{\ref{YY}{c}} \label{YYc}
        \end{align}

        The optical mode and the mechanical mode has onsite energy $\omega_{c}$ and $\omega_{m}$, respectively. Note that we didn't include the driving term $\hat{H}_{\rm drive}=\sum_{n=1}^{n=N} (i \sqrt{\kappa_{ex}}\alpha_{in} \hat{a}^{\dagger}_{n} e^{-i \omega_{L}t}+{\rm H.c.}) $ into the Hamiltonian here because it only provides a steady optical field here . What we care aboute here is the small fluctuation over this steady optical field. Using the rotation frame transformation $H_{\rm rot}=UHU^{\dagger}-i\hbar U \frac{\partial U^{\dagger}}{\partial t}$ where $U=\exp[\sum_{n}^{n=N}i\omega_{L} t \hat{a}^{\dagger}_{n}\hat{a}_{n}]$ and following the standard linearization procedure $\hat{a}_{n}=\bar{\alpha}_{n}+\delta \hat{a}_{n} $ , $\hat{b}_{n}=\bar{\beta}_{n}+\delta \hat{b}_{n}$ one would have the lattice Hamiltonian
        \begin{equation}
            \begin{aligned}
                \hat{H}=\sum_{n=1}^{n=N} &\bigg [ (-\Delta-i \kappa) \hat{a}^{\dagger}_{n}\hat{a}_{n}+(\omega_{m}-i \Gamma)\hat{b}^{\dagger}_{n}\hat{b}_{n}\\
                &+G(\hat{a}^{\dagger}_{n}+\hat{a}_{n})(\hat{b}^{\dagger}_{n-1}+\hat{b}_{n-1})
                +G(\hat{a}^{\dagger}_{n}+\hat{a}_{n})(\hat{b}^{\dagger}_{n}+\hat{b}_{n})\\ &+(t-\frac{\gamma}{2})\hat{a}^{\dagger}_{n+1}\hat{a}_{n}+(t+\frac{\gamma}{2})\hat{a}^{\dagger}_{n}\hat{a}_{n+1} \bigg ]
            \end{aligned}
        \end{equation}
        where $\Delta=\omega_{L}-\omega_{c}$ is the detuning of the driving laser and $G=\bar{\alpha} g$ is relative optomechanical interaction.Note that we have assumed   $\bar{\alpha}_{n}=\bar{\alpha},\bar{\beta}_{n}=\bar{\beta}  $ for all $n$ which is reasonable for the system has translation invariance in the bulk for large size.  And we have replace $\delta \hat{a}_{n} (\delta \hat{b}_{n})$ with $ \hat{a}_{n} ( \hat{b}_{n})$ for simplicity. Our paper foucsed on the ``beam splitte'' regime where $\Delta\approx -\omega_{m}$ and the lattice can be considered as a system composed of two bosonic subsystems with the same onsite energy which can interchange quanta. In this regime the energy non-conserving term $\hat{a}^{\dagger}\hat{b}^{\dagger} $ and $\hat{a} \hat{b}$ can be safely dropped. Since  $-\Delta\approx \omega_{m}$ we can omit the onsite energy because it is just a constant energy shift or we can perform the rotation frame transformation $U^{'}=\exp[\sum_{n}^{n=N}i\omega_{m} t (\hat{a}^{\dagger}_{n}\hat{a}_{n}+\hat{b}^{\dagger}_{n}\hat{b}_{n})]$ again to eliminate it. So the Hamiltonian we discuss in the text below is 
        \begin{equation}
            \begin{aligned}
                \hat{H}=\sum_{n=1}^{n=N} &\bigg [ -i \kappa \hat{a}^{\dagger}_{n}\hat{a}_{n}-i \Gamma\hat{b}^{\dagger}_{n}\hat{b}_{n}+G(\hat{a}^{\dagger}_{n}\hat{b}_{n-1}+\hat{b}^{\dagger}_{n-1}\hat{a}_{n})
                \\ & +G(\hat{a}^{\dagger}_{n}\hat{b}_{n}+\hat{b}^{\dagger}_{n}\hat{a}_{n})+(t-\frac{\gamma}{2})\hat{a}^{\dagger}_{n+1}\hat{a}_{n}+(t+\frac{\gamma}{2})\hat{a}^{\dagger}_{n}\hat{a}_{n+1} \bigg ] \label{ham}
            \end{aligned}
        \end{equation}

\subsection{Generalized Brillouin zone and energy spectrum of the system}
  The eigenstates and eigenvalues  can be solved numrically for the finite non-Hermitian optomechanical lattice. While in order to understand the features (e.g. the topology , the non-Hermitian skin effect) capturing the system more accurately, it's still worth investigating the system analytically. Since the analytical method for non-Hermitian system, i.e. non-Bloch band theory\cite{yokomizo2019non,kawabata2020non,yao2018edge,yao2018non,yang2020non,zhang2020correspondence,longhi2020non}, has been well established , we will follow its standard way and give the generalized Brillouin zone (GBZ) which influences many physical properties  of the system.
  
  The eigenstate of the system in real space can be written as $| \psi \rangle =(\psi_{1,A},\psi_{1,B},\psi_{2,A},\psi_{2,B},\cdots )^{T} $  and the eigenfunction $H| \psi \rangle=E| \psi \rangle$ can be divided into a set of cascaded equations
  \begin{equation}
    \left\{
    \begin{aligned}
   E \psi_{n,A}&=-i \kappa \psi_{n,A}+G\psi_{n-1,B}+G\psi_{n,B}+(t-\frac{\gamma}{2})\psi_{n-1,A}
    \\ & +(t+\frac{\gamma}{2})\psi_{n+1,A} 
      \\ \vspace*{4cm}
    E  \psi_{n,B}&=-i \Gamma \psi_{n,B}+G\psi_{n+1,A}+G\psi_{n,A}\\
    \end{aligned}
    \right.
    \end{equation}
    
    Take the ansatz $\psi_{n,\mu}=\sum_{j} \phi^{(j)}_{n,\mu},(\mu=A,B)$, where $(\phi^{(j)}_{n,A},\phi^{(j)}_{n,B})=(\beta_{j})^{n}(\phi^{(j)}_{A},\phi^{(j)}_{B})$ , one have 
    \begin{equation}
        \left\{
        \begin{aligned}
        &E \phi_{A}=-i \kappa \phi_{A}+G \beta^{-1} \phi_{B}+G\phi_{B}+(t-\frac{\gamma}{2})\beta^{-1}\phi_{A}+(t+\frac{\gamma}{2})\beta \phi_{A} \\
        &E  \phi_{B}=-i \Gamma \phi_{B}+G \beta \phi_{A}+G\phi_{A}\\
        \end{aligned}
        \right.
        \end{equation}
    So the generalized Bloch Hamiltonian $H(\beta)$ is defined as
    \begin{equation}
        H(\beta)=\\  
          \left(
          \begin{array}{cc}
              -i\kappa+(t-\frac{\gamma}{2})\beta^{-1}+(t+\frac{\gamma}{2})\beta &\hspace{1cm} G+G \beta^{-1}\\
              G+G \beta & -i \Gamma \label{nonblochh}
          \end{array}
          \right)
          \end{equation}
    In order to get the eigenvalues of the system we have to solve the non-Bloch equation $H(\beta)| \psi \rangle=E| \psi \rangle$ which leads to $\rm det(H(\beta)-E\mathbb{I} )=0$, i.e. a quadratic equation $f(\beta,E)$ for both $E$ and $\beta$,
    \begin{equation}
        \begin{aligned}
            f(\beta,E)=E^{2}+a(\beta)E+b(\beta)=0 \label{ee}
        \end{aligned}
    \end{equation}
where
    \begin{equation}
        \left\{
        \begin{aligned}
            &a(\beta)=[-(t+\frac{\gamma}{2})\beta+i(\Gamma+\kappa)-(t-\frac{\gamma}{2})\beta^{-1}]\\
            &b(\beta)=(-i\Gamma(t+\frac{\gamma}{2})-G^{2})\beta-(\Gamma \kappa+2G^{2})-(i\Gamma(t-\frac{\gamma}{2})+G^{2})\beta^{-1}
        \end{aligned}
        \right.
    \end{equation}
    
    To have a  continuous energy band one must have $\vert \beta_{1}\vert=\vert \beta_{2}\vert $ for Eq.\ref{ee} and all  $\beta$ satisfy this condition will form the GBZ we want\cite{yokomizo2019non,yu2021generalized}. That is to say, if we have $\beta$ satisfies $f(\beta,E)=0$ , there will be another $\tilde{\beta}=\beta e^{i\theta} $ satisfying $f(\beta e^{i\theta},E)=0$. Using the method of resultant one can eliminate $E$ by getting $R_{E}(\beta,e^{i\theta})=0$\cite{yang2020non}. For any $\theta\in [0,2\pi]$ the resultant equation will give solution $\beta$ that satisfies $f(\beta,E)=f(\beta e^{i\theta},E)=0$ . By varying $\theta$ one can get a set of $\beta$ which constitutes   the auxiliary generalized Brillouin zone (aGBZ)   . For  equation $f(\beta,E)=0$  with higher order $2M$ ($M>1$) for $\beta$ will have roots ${\beta_{1},\beta_{2},\cdots,\beta_{2M}}$ listed by the modulus ($\vert \beta_{1}\vert \leq \vert \beta_{2}| \leq \cdots \leq \vert \beta_{2M}\vert $), one have to verifying whether $\beta_{0}$ and $\beta_{0}e^{i\theta_{0}}$ given by resultant equation $R_{E}(\beta,e^{i\theta})=0 $ is the middle two root $\beta_{M}$ and $\beta_{M+1}$. That's so so-called picking up the GBZ from the aGBZ, while we don't have to do this because we have only two roots here and they are apparantly the middle two roots. So the aGBZ we get is actually the GBZ we want.

    Typical parameters for optomechanical system can be found in previous experimental work\cite{aspelmeyer2014cavity}. Usually we have optical and mechanical damping rates in the range $ 10^{5}\sim  10^{9} \rm Hz$ and $1\sim 10^{4} \rm Hz$. And the optomechanical interaction constant ranges from $1 \rm Hz$ to $10^{5} \rm Hz$ . The optical mode coupling constant $t$ and nonreciprocity $\gamma$ can be tuned on a large scale by adjusting the waveguide that connects them. In our paper, we set optical mode coupling constant $t$ as the unit of the energy, and all other parameters have their value relative to it. \\
    \begin{figure*}
        \centering
        \includegraphics[width=\linewidth]{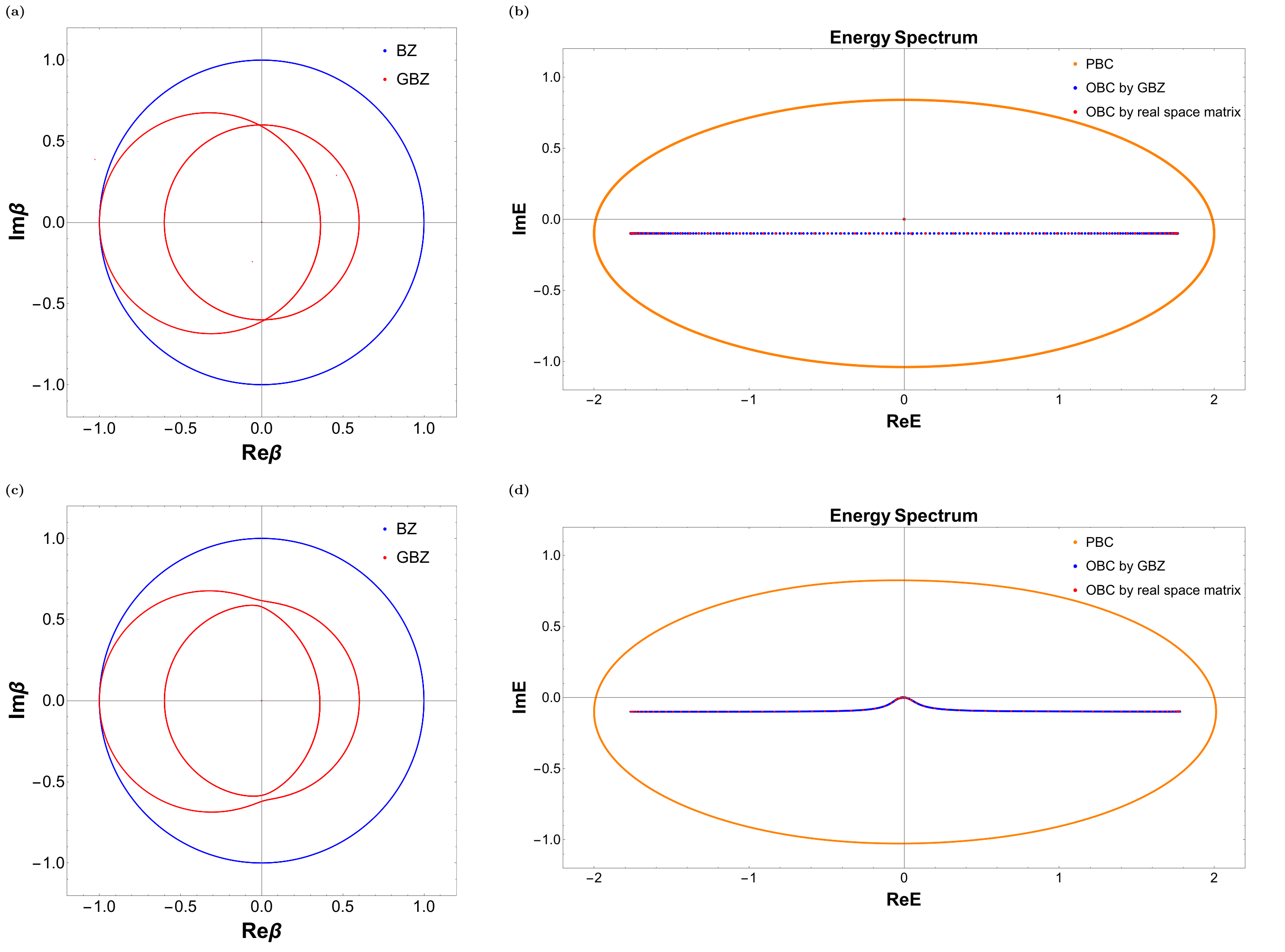}
        \caption{Brillouin zone, generalized Brillouin zone and energy spectrum with different parameters. (a)(b)$t=1,\kappa=0.1,G=0.0008,\Gamma=10^{-6},\gamma=0.94$ (c)(d) $t=1,\kappa=0.1,G=0.08,\Gamma=10^{-6},\gamma=0.94$ . In (a) and (c), the blue unit circle represents the BZ while red circles represent the GBZ. In (b) and (d), the red dots are eigenvalues calculated by solving eigen equation of a 60-site lattice system, while the blue dots (lines) are OBC eigenvalues given by the GBZ in (a) and (c), respectively. The orange  loop  represents the PBC energy spectrum. }
        \label{gbz}
    \end{figure*}
      We have shown two examples of how GBZ affects the energy spectrum of the system in Fig.\ref{gbz}. In Fig.\ref{gbz} (a) and (c), we have shown the BZ and the GBZ for different parameters, i.e., $t=1,\kappa=0.1,G=0.0008,\Gamma=10^{-6},\gamma=0.94$ for Fig.\ref{gbz} (a) and $t=1,\kappa=0.1,G=0.08,\Gamma=10^{-6},\gamma=0.94$ for Fig.\ref{gbz} (c). Fig.\ref{gbz} (b) and  (d) are the energy spectrum for Fig.\ref{gbz} (a) and  (c), respectively. We can see two red intersecting circles constitute the GBZ in Fig. \ref{gbz} (a) which is reasonable for our two band model. Each red circle is the sub-GBZ which forms the energy band of its own. Specifically, the left bigger red circle is the sub-GBZ that forms the mechanical-like energy band focused around the origin of coordinates. This is easy to understand because the mechanical mode has a decay rate much smaller than the optical mode, so the imaginary part of its energy should be close to zero.  The right smaller circle in Fig.\ref{gbz} (a) is the sub-GBZ responsible for forming the flat optical-like energy band with the imaginary part around $-0.1$ and the real part ranging from $-1.76$ to $1.76$. We can see two different ways to calculate the energy spectrum, i.e. solving the real space Hamiltonian matrix eigen equation or using the GBZ to give the energy spectrum, leading to the same result. In Fig.\ref{gbz} (b) and (c), the red points represent  the brute force calculation of a 60-site real space Hamiltonian and the blue dots represent the GBZ way. By denoting our system 60-site we mean our system are composed of 60 optical modes and 60 mechanical modes. Although both methods can give the energy spectrum, the GBZ way can give more calculational points than the real space Hamiltonian way during the same calculational time, thus revealing the energy spetrum more accurately. In Fig.\ref{gbz} (c) and (d) the optomechanical interaction $G$ is stronger, so the mechanical mode and the optical mode interacts with each other strongly and they form the new polariton modes. This can be shown both in the GBZ and the energy spectrum. In Fig.\ref{gbz} (c), the GBZ becomes two red circle lines with the smaller one surrounded by the bigger one. One can check that the left part of the bigger circle and the right part of the smaller circle is the sub-GBZ for the mechanical-like polariton mode. However,   the right part of the bigger circle and the left part of the smaller circle is the sub-GBZ for the optical-like polariton mode. So the sub-GBZ for different energy band mingles with each other because of  a stronger optomechanical interaction $G$ . We  can also see this feature in Fig.\ref{gbz} (d), where the optical-like polariton mode transforms from the flat band to a parabola-like band  . By comparing Fig.\ref{gbz} (b) and  (d), we can also see the imaginary gap for two bands shrinks from $0.1$ to $0$ . Since the imaginary part of the eigen energy corresponds to the decay of the eigenstate, we can conclude that with the  increase  of the optomechanical interaction strength $G$ the eigenstates become less dissipative. We should also note that whether in Fig.\ref{gbz} (a) and (c) the GBZ is surrounded by the unit circle BZ, which means every $\beta$ in the GBZ has a modulus less than 1. And this is the reason why we will have non-Hermitian skin effect in our model. Besides the open boundary condition (OBC) energy spectrum we discussed  above, we have also calculated the periodical boundary condition (PBC) energy spectrum in Fig.\ref{gbz} (b) and (d), one can see that PBC spectrum is a orange closed loop that surrouds the blue OBC pectrum.

\subsection{Non-Hermitian skin effect of the  system}
\begin{figure}
    \includegraphics[width=\linewidth]{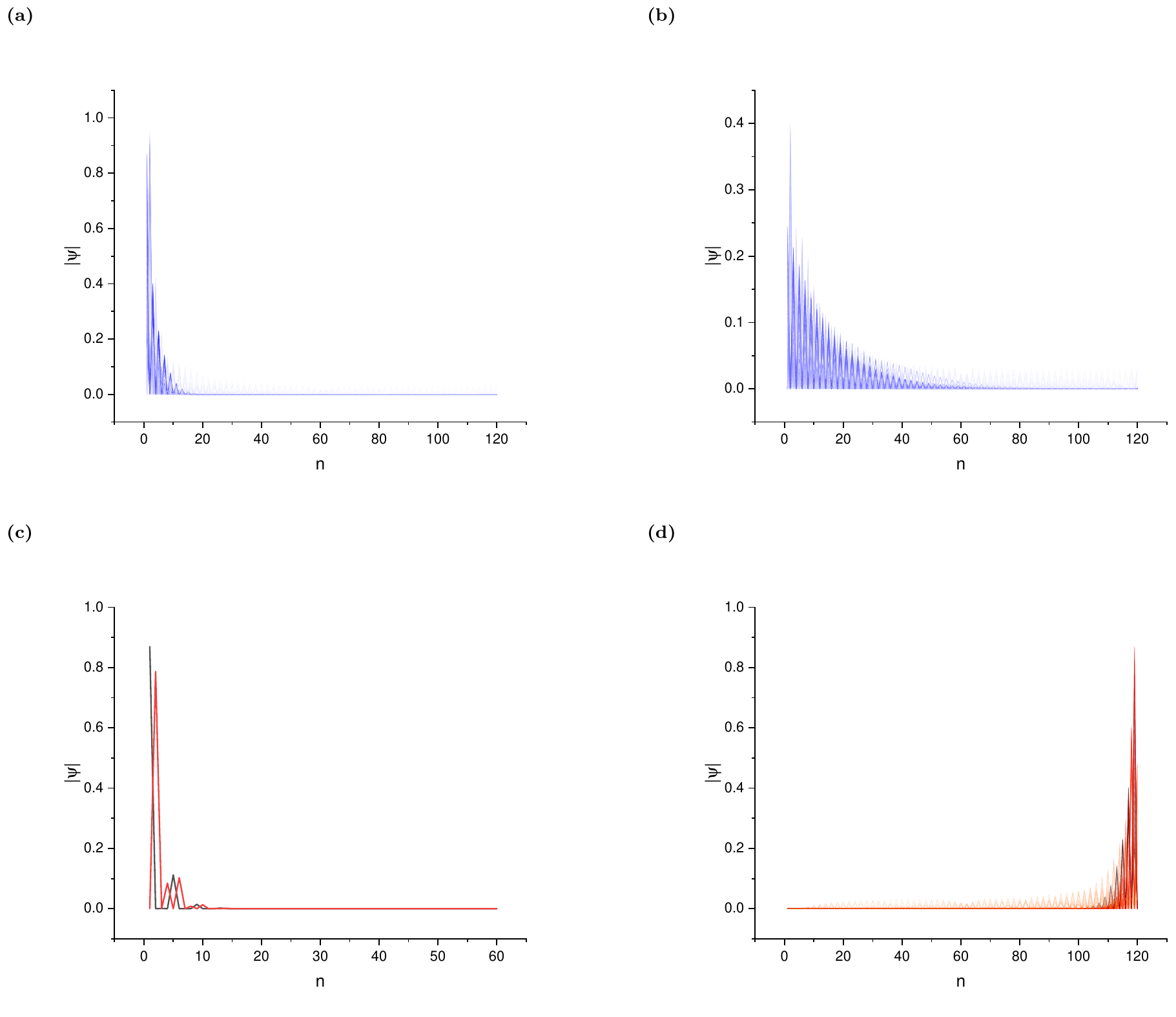}
    \caption{Skin effect of our 60-site lattice system. From (a) to (d), all parameters are the same except for  nonreciprocity $\gamma$. They are $t=1,\kappa=0.1,G=0.0008,\Gamma=10^{-6}$. In (a) and (c), $\gamma=0.94$. In (b) $\gamma=0.14$ and in (d) $\gamma=-0.94$. In (a),(b) and (d), different color lines represent different eigenstate. In (c), the black line represents the optical mode while the red  line represents the mechanical mode.   }  
    \label{skinpic}
\end{figure}

   By looking at Fig.\ref{gbz}, one would find our GBZ could be constituted by $\beta$ with modulus less than $1$ under the choice of system parameters, so our system might have non-Hermtian skin effect\cite{okuma2020topological,yao2018edge}. This is shown in Fig.\ref{skinpic}. By comparing Fig.\ref{skinpic} (a) and (d), one can see that the nonreciprocity $\gamma$ will enhance the skin effect since (a) has a shorter decay length than (d). By comparing Fig.\ref{skinpic} (a) and (d), one can see that the symbol of $\gamma$ will affect the side that skin effect appears.  In Fig.\ref{skinpic} (c) we show two specific eigenstate, where the black line represents the optical skin mode while the red line represents the mechanical skin mode.  

\section{Green function of the system \label{section three}}
 Our main goal is to discuss the transmission property of our non-Hermitian optomechanical lattice, so it's worth calculating the Green function of the system.  Similarly , there are two ways to calculate the Green function of the system, one is solving the real space matrix and the other one is using the formulas of Green function in non-Bloch  band theory  given by Xue\cite{xue2021simple}.
 
\subsection{Solving the real space Green function matrix}
Our system is an open quantum system described by the quantum master equation 
\begin{equation}
    \begin{aligned}
        \dot{\rho}(t)=-i[H_{0},\rho]+\sum_{\mu}(L_{\mu}\rho L^{\dagger}_{\mu}-\frac{1}{2}\left\{L^{\dagger}_{\mu}L_{\mu},\rho\right\}) 
    \end{aligned}
\end{equation}

where $H_{0}$ reads 
\begin{equation}
    \begin{aligned}
        H_{0}=\sum_{n} & \bigg [t(a^{\dagger}_{n+1}a_{n}+a^{\dagger}_{n}a_{n+1})+G(a^{\dagger}_{n}b_{n-1}+b^{\dagger}_{n-1}a_{n})\\
        & +G(a^{\dagger}_{n}b_{n}+b^{\dagger}_{n}a_{n})+(i \sqrt{\kappa_{ex}} \bar{\alpha}_{in}(t)a_{n}+\rm h.c. )\bigg ]
    \end{aligned}
\end{equation}

   By setting $\left\{ L_{\mu}\right\}=\left\{ \sqrt{2\Gamma}b_{i},\sqrt{\gamma}(a_{i}-ia_{i+1}),\sqrt{\kappa_{ex}}a_{i}\right\}$ ,one can show the expectations of the operator $a_{i}$ and $b_{i}$ , i.e. $<a_{i}(t)>=Tr[a_{i}{\rho}(t)]$ and $<b_{i}(t)>=Tr[b_{i}{\rho}(t)]$ , are governed by the Langevin equations

   \begin{equation}
    \left\{
    \begin{aligned}
    <\dot{a}_{i}(t)>=&\frac{-\kappa_{ex}-2\gamma}{2}<a_{i}>-i[(t+\frac{\gamma}{2})<a_{i-1}>\\ & +(t-\frac{\gamma}{2})<a_{i+1}>]-iG(<b_{i-1}>+<b_{i}>)\\ &-\sqrt{\kappa_{ex}}\bar{\alpha}_{in}(t)\\
    <\dot{b}_{i}(t)>&=-\Gamma  <b_{i}(t)>-iG(<a_{i+1}>+<a_{i}>)\\
    \end{aligned}
    \right.
    \end{equation}

    By denoting $(\psi_{1},\psi_{2},\psi_{3},\cdots,\psi_{2n-1},\psi_{2n},\cdots)=(<a_{1}>,<b_{1}>,<a_{2}>, \cdots ,<a_{n}>,<b_{n}>\cdots)$, one would have the matrix equation:
    \begin{equation}
        \begin{aligned}
            \dot{\psi}_{i}=-i \sum_{j} H_{ij} \psi_{j}- \sqrt{\kappa_{ex}} \psi_{in,i}
        \end{aligned}
    \end{equation}
    where $H$ is exactly the Hamiltonian in  Eq.\ref{ham} with $\frac{\kappa_{ex}+2\gamma}{2}=\kappa$ and $\psi_{in}=(\bar{\alpha}_{in},0,\bar{\alpha}_{in},0,\cdots,\bar{\alpha}_{in},0,\cdots)$. If we are intereted in the transmission property of the frequency $\omega$ component , by performing the Fourier tranformation for both ends, we will get 
    \begin{equation}
        \begin{aligned}
            \vec{\psi}(\omega)=-i\sqrt{\kappa_{ex}} G(\omega) \vec{\psi}_{in}(\omega)
        \end{aligned}
    \end{equation}
    where $\vec{\psi}(\omega)=(\psi_{1}(\omega),\psi_{2}(\omega),\cdots,\psi_{2n-1}(\omega),\psi_{2n}(\omega),\cdots)$ and similar for the definition of $\vec{\psi}_{in}(\omega)$. $G(\omega)=\frac{1}{\omega-H}$ is the Green function of our optomechanical lattice with uniform driving field for each optical mode.  Using the input-output relathion $\psi_{j,out}=\psi_{j,in}+\sqrt{\kappa_{ex}}\psi_{j}$ we can immediately get $\psi_{j,out}=(\mathbb{I} -i\kappa_{ex}G)_{jk}\psi_{k,in}$. And the scattering matrix is naturally defined as $S_{jk}(\omega)=(\mathbb{I} -i\kappa_{ex}G(\omega))_{jk}$.
      We note that we didn't consider the effect of the probe laser here. In fact, the introduction of the probe laser can be viewed as an effective imaginary potential\cite{mcdonald2018phase}
      \begin{equation}
        \begin{aligned}
            V_{ij}=-i \frac{\kappa_{p,j}}{2} \delta_{ij}
        \end{aligned}
    \end{equation}
    where $\kappa_{p,j}$ is the loss introduced by probe laser at $j$ site. Using the Dyson's equation one can get the Green equation with probe laser's contribution $\tilde{G}(\omega)$
    \begin{equation}
        \begin{aligned}
            \tilde{G}(\omega)=G(\omega)+ G(\omega)V \tilde{G}(\omega)
        \end{aligned}
    \end{equation}
    And similarly the scattering matrix element in probe laser's frequency domain is
    
    \begin{equation}
        \begin{aligned}
            S_{jk}(\omega_{p})=\frac{\psi_{j,out}}{\psi_{k,in}}=(\mathbb{I} -i\sqrt{\kappa_{p,j}}\sqrt{\kappa_{p,k}}\tilde{G}(\omega_{p}))_{jk} \label{greenprobe}
        \end{aligned}
    \end{equation}
\subsection{Green function given by non-Bloch band theory  }
  In this subsection, we will show how to use the non-Bloch band theory to calculate the Green function of our system and compare it with the brute force calculation of the real space calculation. According to Xue\cite{xue2021simple}, the formulas for Green function of a non-Hermitian system is given by

  \begin{equation}
    \begin{aligned}
        G_{ij}=\int_{GBZ} \frac{ d\beta}{2\pi i \beta} \frac{\beta^{i-j}}{\omega-h(\beta)} \label{greenformulas}
    \end{aligned}
\end{equation}
\begin{figure*}
    \centering
    \includegraphics[width=\linewidth]{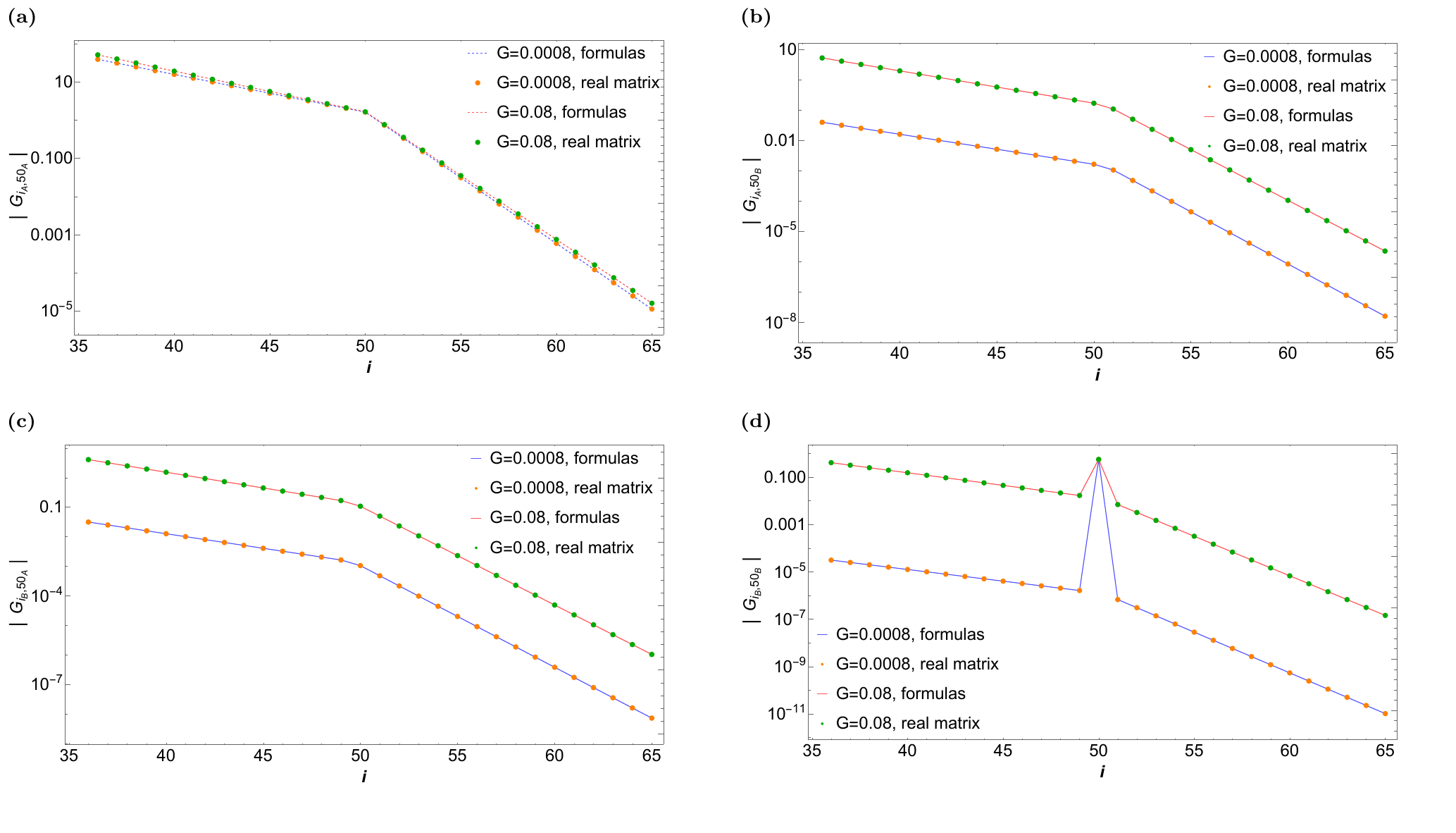}
    \caption{Green function calculated by two ways for a 100-site system . From (a) to (d), the parameters are $t=1,\kappa=0.1,\Gamma=10^{-6},\gamma=0.94,\omega=1.8$ . The orange and green dots are Green function calculated by solving the real space matrix with 100 sites. The blue and red line are calculated by the formulas for Green function in non-Bloch band theory given by Eq.\ref{greenformulas}. The blue line and the orange dots ($G=0.0008$) fit very well, so do the red line and the green  dots ($G=0.08$) .  One can see that two calculational ways have almost the same result .   }
    \label{greenpic}
\end{figure*}
 Since $h(\beta)$ is $2\times 2$ matrix, $G_{ij}$ is also a $2 \times 2$ matrix 
  
 \begin{equation}
    G_{ij}=\\  
      \left(
      \begin{array}{cc}
          G_{i_{A}j_{A}} &\hspace*{0.5cm}G_{i_{A}j_{B}}\\
          G_{i_{B}j_{A}} &\hspace*{0.5cm} G_{i_{B}j_{B}}
      \end{array}
      \right)
      \end{equation}
where 
\begin{equation}
    \begin{aligned}
        G_{i_{\mu}j_{\nu}}=\int_{GBZ} \frac{ d\beta}{2\pi i \beta} \beta^{i-j} g_{i_{\mu}j_{\nu}}(\beta)     \label{gmunv}  
    \end{aligned}
\end{equation}
 $(\mu,\nu=A,B)$. For example, $G_{i_{A}j_{B}}$represents the process of a mechanical mode in site $j$ transporting to a optical mode in site $i$ and similar for other matrix elements' definition. We show the calculation process for $G_{i_{A}j_{A}}$  here.
  By solving the inverse of the matrix $\omega-h(\beta)$ one can show that $g_{i_{A}j_{A}}(\beta)=\frac{\beta}{A\beta^{2}+B\beta+C}$ with $A=\frac{2iG^{2}+(\Gamma-i\omega)(-2t-\gamma)}{2(\Gamma-i\omega)}$ . Substituting  $g_{i_{A}j_{A}}(\beta)$ into Eq.\ref{gmunv} we will get

  \begin{equation}
    \begin{aligned}
        G_{i_{A}j_{A}}(\omega)&=\int_{GBZ} \frac{ d\beta}{2\pi i } \frac{\beta^{i-j}}{A\beta^{2}+B\beta+C}  \\
        &=\int_{GBZ} \frac{ d\beta}{2\pi i A} \frac{\beta^{i-j}}{(\beta-\beta_{1})(\beta-\beta_{2})}
    \end{aligned}
\end{equation}

where $\beta_{1}$ and $\beta_{2}$ are two roots that satisfy $\omega \mathbb{I} -h(\beta)=0$. We denote that $\vert \beta_{1} \vert< \vert \beta_{2} \vert$ , so the residue theorem shows that  
  \begin{equation}
    \begin{aligned}
        &G_{i_{A}j_{A}}(\omega)=\frac{\beta_{1}^{i-j}}{A(\beta_{1}-\beta_{2})} (i\geqq j) \\
        &G_{i_{A}j_{A}}(\omega)=\frac{-\beta_{2}^{i-j}}{A(\beta_{2}-\beta_{1})} (i<j) \label{greenamp}
    \end{aligned}
\end{equation}

\section{Transmission property of the system\label{section four}}
\subsection{Amplification of the system}
  We have calculated the Green function of the system in two different ways. As shown in Fig.\ref{greenpic}, one can see that two different calculational ways give almost the same result for a 100-site system. Each subfigure in Fig.\ref{greenpic} has two lines, one represents weak coupling region $G=0.0008$ and the other represents strong coupling region $G=0.08$. Other parameters are the same:$t=1,\kappa=0.1,\Gamma=10^{-6},\gamma=0.94,\omega=1.8$. In Fig.\ref{greenpic} (a) one can see that the optical mode favors transporting from right to left rather than the opposite direction, which serves as an evidence for non-Hermitian skin effect. From Fig.\ref{greenpic} (d) we can see there is a peak at $G_{50_{B},50_{B}}$ . A closer inspection tells us that this peak becomes smaller with the increase of the optomechanical interaction $G$. This means under the parameter $G=0.0008$ the phonon is more likely to be localized rather than hopping to the neighboring optical mode, which comes as a natrual result of weak optomechanical interaction. By increasing the optomechanical interaction $G$ , e.g. $G=0.08$,  one can find this peak will become less  pronounced. Except for the amplification between optical modes, we can also also achieve the amplification from the optical mode on the right side to the mechanical mode on the left side . For example, one can have $\vert G_{1_{B},50_{A}} \vert = 58.58 $   which is not shown in the blue line in Fig.\ref{greenpic}(c). Similarly, we can have amplification from mechanical mode to optical mode and amplification from mechanical mode to mechanical mode by choosing proper $\omega$ with other parameters unchanged . The direction amplification is always from right to left due to non-Hermitian skin effect, so that our system can be considered as a single pass filter.
  \begin{figure*}
    \centering
    \includegraphics[width=\linewidth]{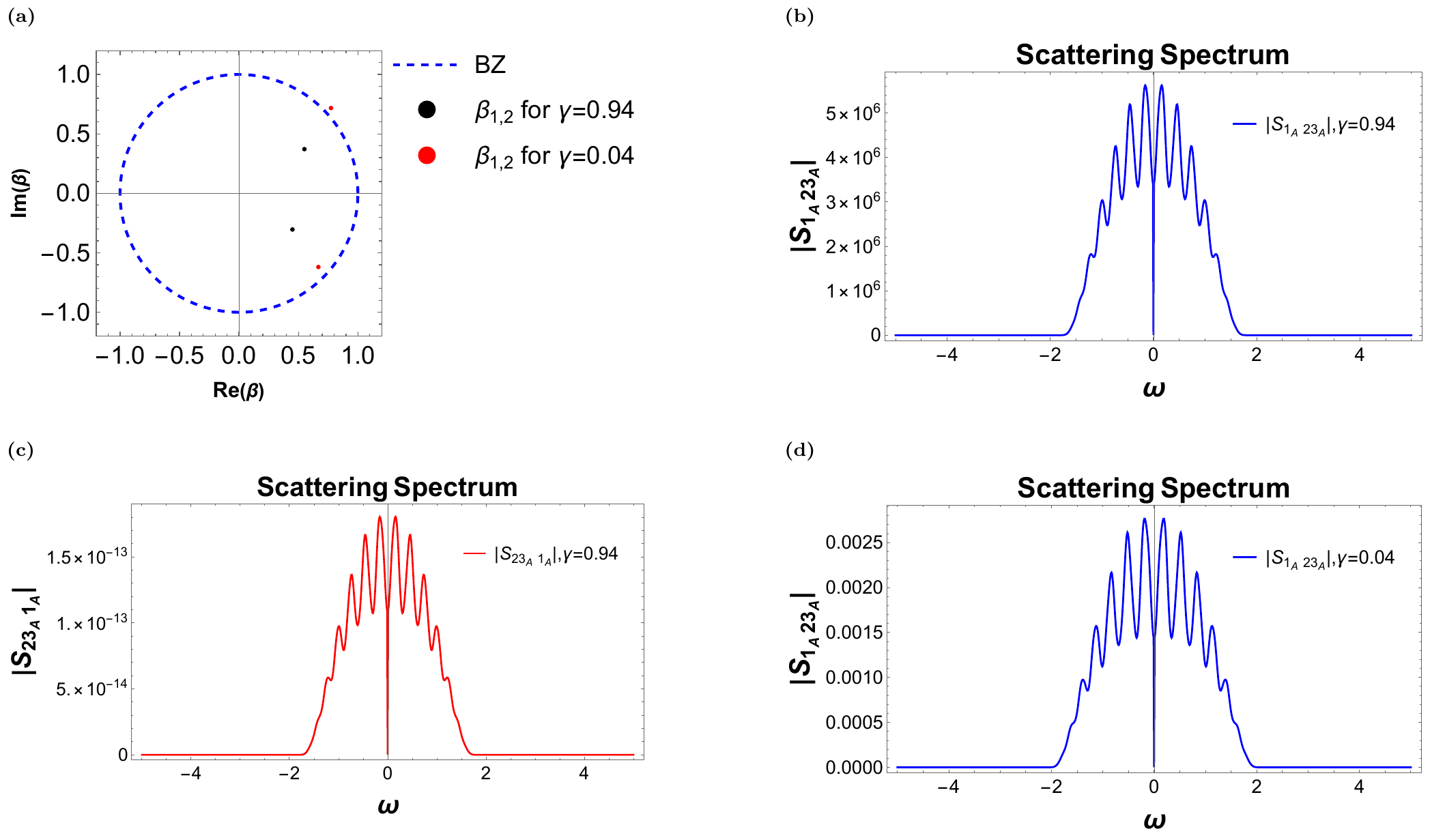}
    \caption{Scattering spectrum of a 40-site lattice optomechanical system. We have added two probe laser at site 1 and site 23 and investigate the transmission property between them. In (b),(c) and (d), all parameters are the same except for  nonreciprocity $\gamma$ : $t=1,\kappa=0.1,G=0.0008,\Gamma=10^{-6},\kappa_{p,1}=\kappa_{p,23}=0.1$ .In (b) and (c) $\gamma=0.94$ while in (d) $\gamma=0.04$. Similarly, we have valued $\omega$ as its strength relative to $t$. Since we have performing rotation frame transformation before, $\omega=0$ here means $\omega_{p}=\omega_{a}$.  In (a)  we show the roots for $\omega \mathbb{I} -h(\beta)=0$ with $\omega=1.47$ and their position with respect to the BZ. Other parameters in (a) are: $t=1,\kappa=0.1,G=0.0008,\Gamma=10^{-6}$.  }\label{scatterpic}
\end{figure*}

  Now we discuss the amplification behavior of the system with help of Eq.\ref{greenprobe}. In Fig.\ref{scatterpic}, we have shown the scattering property of a 100-site system  where two probe laser are added on the 1st and the 23rd optical mode. From Fig.\ref{scatterpic} (b) and (c) one can see evident amplification from site 23 to site 1 while no amplification from site 1 to site 23 when $\gamma=0.94$. From Fig.\ref{scatterpic} (d) we can see no amplification even from site 23 to site 1 with $\gamma=0.04$ . This can be explained by  Eq.\ref{greenamp}  . In Fig.\ref{scatterpic}(a) we plot the roots $\beta_{1}, \beta_{2}$ for $\omega \mathbb{I} -h(\beta)=0$. From Fig.\ref{scatterpic}(a) we can see under parameter $\gamma=0.94, \omega=1.47$ even the bigger root $\vert \beta_{2} \vert<1 $, so that we have $G_{i_{A}j_{A}}\gg 1$ for $i\ll j$. While this behavior will vanish when $\gamma=0.04,\omega=1.47$ because we have $\vert \beta_{2} \vert>1 $. This will lead to $G_{i_{A}j_{A}}\ll 1$ for $i\ll j$ even though from an intuitive sight we might think the amplification still exists  since we still have $t+\frac{\gamma}{2}>t-\frac{\gamma}{2}$. Another important feature is that the scattering spectrum is symmetric about  $\omega=0$ and there is a sharp dip at $\omega=0$ no matter there is amplification or not. We note that $\omega=0$ corresponds to condition $\omega_{p}=\omega_{a}$  here and we will show the origin of  the dip at $\omega=0$ in our next subsection.
  
  Another important question is to determine the region of $\omega$ that supports amplification. From Fig.\ref{scatterpic} (b) we can see this region is roughly $-2<\omega<2$. This can be seen from our PBC spectrum in Fig.\ref{gbz} (b) and (c). Suppose we have a vertical line which intersects with the real axis at $(\omega,0)$  in Fig.\ref{gbz} (b) and (c) . When $\omega\rightarrow \pm \infty$ there are no interaction point between this vertical line and the PBC spectrum. One can prove that the smaller root for $\omega \mathbb{I} -h(\beta)=0$, i.e. $\beta_{1}$ stays inside the  PBC spectrum while the bigger root $\beta_{2}$ stays out of the PBC spectrum as long as this vertical line has no intersection point with the PBC spectrum\cite{yu2021generalized}. By moving this vertical line from $\pm \infty$ to the center , when the vertical line has intersection points with the PBC spectrum, the outer $\beta_{2}$ will move into the BZ since the BZ generates the PBC spectrum. And that's the moment when amplification starts  existing. Since the first intersection point between the vertical line and the PBC spectrum has real part about $\pm 2$, the region $(-2,2)$ is region that supports amplification.
  
\subsection{Optomechanically induced transparency in our system}
\begin{figure*}
    \centering
    \includegraphics[width=\linewidth]{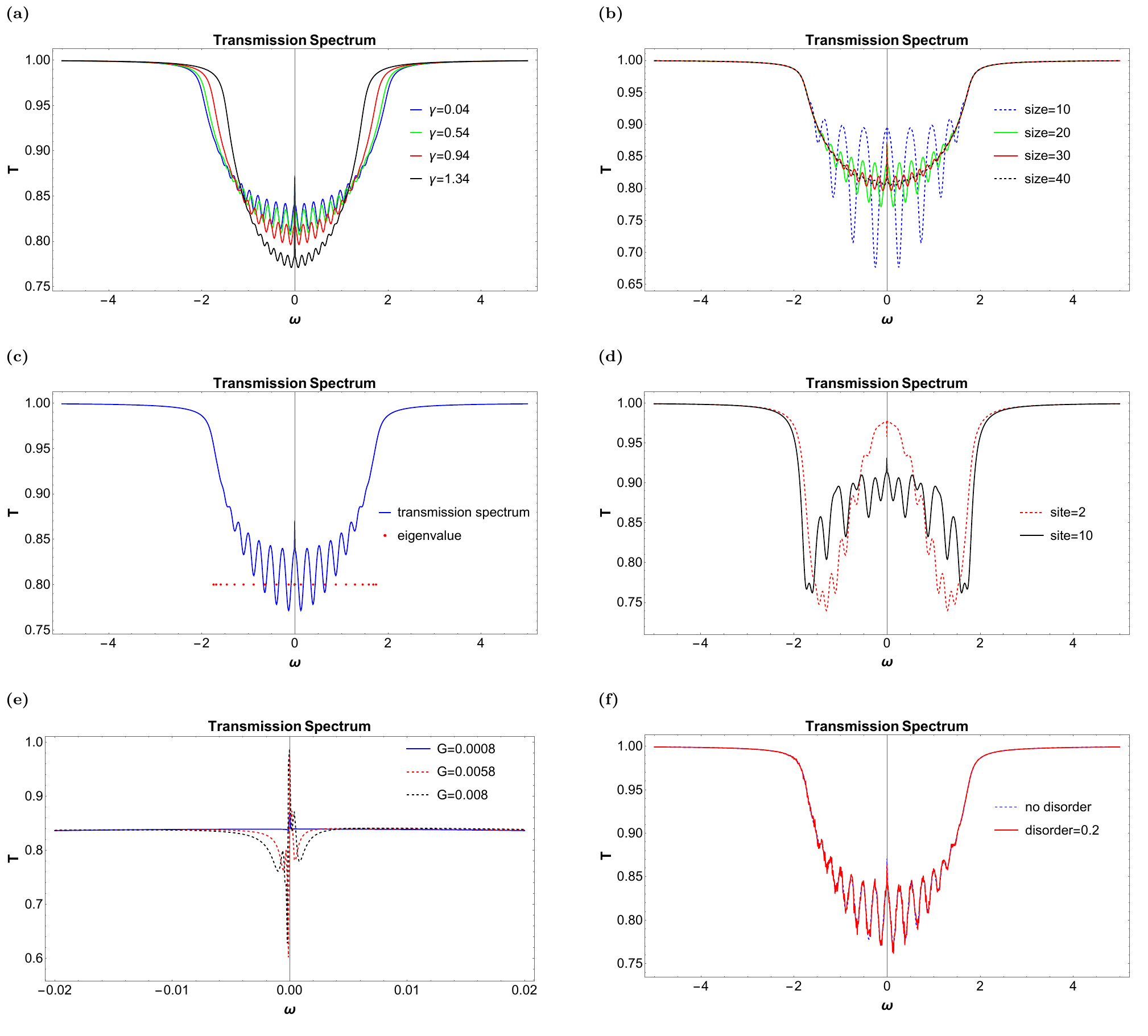}
    \caption{Optomechanically induced transparency of the first optical site. System parameters: (a) OMIT of a 30-site system. $t=1,\kappa=0.1,G=0.0008,\Gamma=10^{-6},\kappa_{p,1}=0.1$. By increasing the nonreciprocity $\gamma$, one can see that the OMIT peak is increasing.  (b) OMIT effect with different system size: $\gamma=0.94,t=1,\kappa=0.1,G=0.0008,\Gamma=10^{-6},\kappa_{p,1}=0.1$. One can see the realization of OMIA is possible by varing the size of the sytem.  (c) The relation ship between the eigenvalues of a 20-site system and the onsite transmission spectrum, prarameters are $\gamma=0.94,t=1,\kappa=0.1,G=0.0008,\Gamma=10^{-6},\kappa_{p,1}=0.1$. The red points represents the real part of the eigenvalues . (d) The  bulk onsite transmission property of a 20-site system with parameters $\gamma=0.94,t=1,\kappa=0.1,G=0.0008,\Gamma=10^{-6}$.  We added probe laser at site 2 (site 10) with $\kappa_{p,2}=0.1(\kappa_{p,10}=0.1)$ for red dashed line (black solid line).  (e) A zoom-in picture of the OMIT peak of a 20-site system. Other parameters are: $t=1,\kappa=0.1,\Gamma=10^{-6},\kappa_{p,1}=0.1,\gamma=0.94$ (f) OMIT with disorder. Other parameters are: $t=1,\kappa=0.1,G=0.0008,\Gamma=10^{-6},\kappa_{p,1}=0.1,\gamma=0.94$.  }\label{omitpic} 
\end{figure*}
  By adding just one probe laser to some site of our system, e.g. the first left optical mode  of our system, we can investigate the onsite transmission property of the system. From Eq.\ref{greenprobe} we know the onsite transmission for the first optical mode is 
  \begin{equation}
    \begin{aligned}
        T_{11}=\vert \frac{\psi_{1,out}}{\psi_{1,in}}\vert ^{2}=\vert 1-i\kappa_{p,1} \tilde{G}_{1_{A}1_{A}}(\omega)  \vert ^{2} \label{onsiteeq}
    \end{aligned}
\end{equation}
  So the only goal is to calculate the Green function $G_{1_{A}1_{A}}$. Using the method discussed in section \ref{section three} , we get the transmission spectrum shown in Fig.\ref{omitpic} . One can see there are optomechanically induced transparency peaks in Fig.\ref{omitpic} (a), (b) and (c). In Fig.\ref{omitpic} (a), the  transmission spectrum of the first left optical site of  a 30-site system is shown. One can see that with the increase of the nonreciprocity $\gamma$ the OMIT peak will increase under the parameters we choose . So the nonreciprocity might be considered as a method to   modulate the peak of the OMIT.  
  
  Besides, we also investigate the effect of the system's size on the first site transmission property. From Fig.\ref{omitpic}(b) we can see that one can even achieve optomechanically induced amplification (OMIA) in a 10-site system. From Eq.\ref{onsiteeq} we know that this is because $\tilde{G}_{1_{A}1_{A}}(\omega)$ is a complex number, so a different $\rm arg(\tilde{G}_{1_{A}1_{A}}(\omega))$ may turn OMIT into OMIA. This explanation is similar to the famous quantum pathways interference theory in the literature before\cite{weis2010optomechanically}. We can use this  quantum pathways interference theory to explain why an increasing $\gamma$ will lead to an increasing OMIT peak under the parameters in Fig.\ref{omitpic}(a). For a N-site system, there will be N optical eigenstates and N mechanical eigenstates, namely $\left\{\psi_{1,A},\psi_{2,A},\cdots,\psi_{N,A}\right\}$ and $\left\{\psi_{1,B},\psi_{2,B},\cdots,\psi_{N,B}\right\}$. These eigenstates are listed by the order of the modulus of their eigenvalues, i.e. $\left\{\vert E_{1,A}\vert > \vert E_{2,A}\vert > \cdots > \vert E_{N,A}\vert\right\}$ and $\left\{ \vert E_{1,B}\vert>\vert E_{2,B}\vert>\cdots>\vert E_{N,B}\vert \right\}$ where the subscript A represents the optical mode and B represents the mechanical mode. Among all optical eigenstates, there will be an eigenstate $\psi_{N,A}$ which has the smallest eigenvalues $E_{N,A}$. For example, the black line in Fig.\ref{skinpic}(c) is $\psi_{60,A}$ with $\rm \vert Re(E_{60,A})\vert=0.0454$. By increasing the nonreciprocity $\gamma$, one can show that the absolute value of the real part of the smallest optical eigenvalue, i.e. $\rm \vert Re(E_{N,A}) \vert $ , approaches to the zero more closely. So the optical mode $\psi_{N,A}$ has a larger probability to accept the probe laser photon injected into the system with relative frequency $\omega=0$, which finally results in a larger OMIT peak due to a stronger quantum pathways interference. From Fig.\ref{omitpic} (a) and (b) we also find that the transmission spectrum is almost symmetric around $\omega=0$.  A further study on Fig.\ref{omitpic} (b) tells us that the number of the absorption peaks on the left (right) of $\omega=0$ is increasing when the system becomes larger. In fact the absorption peaks on the left (right) of $\omega=0$ are actually the real part of the eigenvalues of the  optical-like modes while the OMIT peak (OMIA dip) centering around $\omega=0$ is connected with the real part of the eigenvalues of the mechanical-like modes. This argument can be testified in Fig.\ref{omitpic} (c), (d) and (e). In Fig.\ref{omitpic} (c), we plot the transmission spectrum of the first optical mode of a 20 site system. At the same time, we calculate the eigenvalues of this 20 site system and mark the real part of the eigenvalues in Fig.\ref{omitpic} (c), i.e. the red dots in  Fig.\ref{omitpic} (c). One can see that they fit quite well with the absorption peaks. To show that the OMIT (OMIA) peak around the center is actucally the mechanical absorption peak, we investigate the detail of the peak with a variable optomechanical interaction $G$. In Fig.\ref{omitpic} (e), one can see that the OMIT peak in weak optomechanical region ($G=0.0008$) splits into various peaks in a stronger optomechanical interaction ($G=0.0058$ or $G=0.008$). Comparing black dashed line and red dashed line in Fig.\ref{omitpic} (e), one would find that the region of split peaks becomes wider with the increase of $G$. This phenomenon is similar to the broadening of the energy band in a tight-banding model. When $G$ is small, the mechanical modes almost decouple from each other. So the mechanical-like eigenvalues form energy levels centering around $\omega=0$, i.e. the OMIT peak . When $G$ is larger, the mechanical modes couple with each other through optomechanical interaction. So the energy levels interact with each other and the mechanical energy band forms consequently. When $G$ is larger, the width of the split region is wider due to a stronger interference between different energy levels.

  From Fig.\ref{omitpic}(b) we also conclude that the optical absorption peaks on the right (left) side of $\omega=0$ become shallower and less pronounced  with system size increases. This is due to the interference of different bulk optical modes. With the increase of the  system size, this many body physics phenomenon becomes more apparant. One can also see this phenomenon by investigating the onsite transmission property of a bulk site,  e.g. the 10th optical mode of a 20-site system as shown as the black dashed line in Fig.\ref{omitpic}(d). One can  find that the optical absorption peaks at the bulk site  become more chaotic than the first site  because the bulk has a stronger interference effect than the edge. Moreover, one can also conclude that changing the site that the probe laser is added to can turn OMIT into OMIA by comparing the red dashed line and black solide line in \ref{omitpic}(d). 

  We finally discuss possibility of the experimental realization of our model. To see this, we consider the effect of inevitable disorders introduced by imperfect fabrication. We define a positive parameter $\epsilon $ which describes the strength of the disorder.
  With this paramter, the Hamiltonian under the effect of disorders is:
  \begin{equation}
    \begin{aligned}
        \hat{H}=\sum_{n=1}^{n=N} &\bigg [ -i (\kappa+\epsilon_{\kappa,n}) \hat{a}^{\dagger}_{n}\hat{a}_{n}-i (\Gamma+\epsilon_{\Gamma,n}) \hat{b}^{\dagger}_{n}\hat{b}_{n}\\ & +(G+\epsilon_{G,n})(\hat{a}^{\dagger}_{n}\hat{b}_{n-1}+\hat{b}^{\dagger}_{n-1}\hat{a}_{n}
         +\hat{a}^{\dagger}_{n}\hat{b}_{n}+\hat{b}^{\dagger}_{n}\hat{a}_{n})\\ &+(t-\frac{\gamma+\epsilon_{\gamma,n}}{2})\hat{a}^{\dagger}_{n+1}\hat{a}_{n}+(t+\frac{\gamma+\epsilon_{\gamma,n}}{2})\hat{a}^{\dagger}_{n}\hat{a}_{n+1} \bigg ] \label{ham}
    \end{aligned}
\end{equation}

  This means that  any parameter $\xi $ in our model can be varied in the region $[\xi-\epsilon_{\xi,n} \xi, \xi+\epsilon_{\xi,n} \xi]$ where $0<\epsilon_{\xi,n}<\epsilon$. As shown in Fig.\ref{omitpic} (f), numerical result shows that the OMIT peak can survive even under a disorder strength as strong as  $\epsilon=0.2$, which shows the robustness of this phenomenon in our model.
  
\section{CONCLUSION}
  We have investigated the energy spectrum and the transmission property of a long non-Hermitian optomechanical lattice . The non-Hermitian property of our model is introduced by  not only the onsite decay but also the nonreciprocal coupling of different sites. To find out the open boundary condition (OBC) spectrum of the system we calculated the generalized Brillouin Zone (GBZ) of the system based on the concept of the resultant.  To investigate the transmission property of the system,  the Green function  is calculated in two ways, i.e. solving the inverse of the real space matrix and using the non-Bloch theory formula. Two different methods give the same result.
  
  With the help of the Green function we investigated the scattering of different sites. The directional amplification from right to left is observed. So our model can serve as a single pathway filter.  The frequency that supports amplification is discussed by considering  the periodical boundary condition energy spectrum of our model . The amplification direction has a close relathionship between the non-Hermitian skin effect. Since our system only supports skin modes on the left side of the lattice, the amplification is always from right to left. As an extension, one can investigate other non-Hermitian Bosonic systems that support bipolar non-Hermitian skin effect.

  Using the Green function, we have also investigated the onsite transmission property of the system. The optomechanically induced transparency (OMIT) is observed in our model. We find that the OMIT peak can be modulated by changing the nonreciprocity $\gamma$ and system size in our model. The absorption peaks are actually corresponding to the real part of the eigenvalues of the system. The mechanical-like mode  eigenvalues are related to   the OMIT peak in the center and the optical-like mode eigenvalues are related to the absorption peak on both sides .  Our model can also be extended to other non-Hermitian Bonsonic systems which have topological features. The possible relathionship between the OMIT and the topology is worth discussing.

  \begin{acknowledgments}
   
    We thank Dr.Yumin Hu and Dr.Jinghui Pi for helpful discussion. This work is supported by the  National Natural Science Foundation
    of China (61727801, 62131002), National Key Research and Development Program of China  (2017YFA0303700), the Key Research and Development  Program of Guangdong province (2018B030325002), Beijing Advanced Innovation Center for Future Chip (ICFC),
    and Tsinghua University Initiative Scientific Research  Program.
    
    \end{acknowledgments}


\begin{thebibliography}{63}%
    \makeatletter
    \providecommand \@ifxundefined [1]{%
     \@ifx{#1\undefined}
    }%
    \providecommand \@ifnum [1]{%
     \ifnum #1\expandafter \@firstoftwo
     \else \expandafter \@secondoftwo
     \fi
    }%
    \providecommand \@ifx [1]{%
     \ifx #1\expandafter \@firstoftwo
     \else \expandafter \@secondoftwo
     \fi
    }%
    \providecommand \natexlab [1]{#1}%
    \providecommand \enquote  [1]{``#1''}%
    \providecommand \bibnamefont  [1]{#1}%
    \providecommand \bibfnamefont [1]{#1}%
    \providecommand \citenamefont [1]{#1}%
    \providecommand \href@noop [0]{\@secondoftwo}%
    \providecommand \href [0]{\begingroup \@sanitize@url \@href}%
    \providecommand \@href[1]{\@@startlink{#1}\@@href}%
    \providecommand \@@href[1]{\endgroup#1\@@endlink}%
    \providecommand \@sanitize@url [0]{\catcode `\\12\catcode `\$12\catcode
      `\&12\catcode `\#12\catcode `\^12\catcode `\_12\catcode `\%12\relax}%
    \providecommand \@@startlink[1]{}%
    \providecommand \@@endlink[0]{}%
    \providecommand \url  [0]{\begingroup\@sanitize@url \@url }%
    \providecommand \@url [1]{\endgroup\@href {#1}{\urlprefix }}%
    \providecommand \urlprefix  [0]{URL }%
    \providecommand \Eprint [0]{\href }%
    \providecommand \doibase [0]{http://dx.doi.org/}%
    \providecommand \selectlanguage [0]{\@gobble}%
    \providecommand \bibinfo  [0]{\@secondoftwo}%
    \providecommand \bibfield  [0]{\@secondoftwo}%
    \providecommand \translation [1]{[#1]}%
    \providecommand \BibitemOpen [0]{}%
    \providecommand \bibitemStop [0]{}%
    \providecommand \bibitemNoStop [0]{.\EOS\space}%
    \providecommand \EOS [0]{\spacefactor3000\relax}%
    \providecommand \BibitemShut  [1]{\csname bibitem#1\endcsname}%
    \let\auto@bib@innerbib\@empty
    \bibitem [{\citenamefont {Ashida}\ \emph {et~al.}(2020)\citenamefont {Ashida},
      \citenamefont {Gong},\ and\ \citenamefont {Ueda}}]{ashida2020non}%
      \BibitemOpen
      \bibfield  {author} {\bibinfo {author} {\bibfnamefont {Y.}~\bibnamefont
      {Ashida}}, \bibinfo {author} {\bibfnamefont {Z.}~\bibnamefont {Gong}}, \ and\
      \bibinfo {author} {\bibfnamefont {M.}~\bibnamefont {Ueda}},\ }\href@noop {}
      {\bibfield  {journal} {\bibinfo  {journal} {Advances in Physics}\ }\textbf
      {\bibinfo {volume} {69}},\ \bibinfo {pages} {249} (\bibinfo {year}
      {2020})}\BibitemShut {NoStop}%
    \bibitem [{\citenamefont {Scully}\ and\ \citenamefont
      {Zubairy}(1999)}]{scully1999quantum}%
      \BibitemOpen
      \bibfield  {author} {\bibinfo {author} {\bibfnamefont {M.~O.}\ \bibnamefont
      {Scully}}\ and\ \bibinfo {author} {\bibfnamefont {M.~S.}\ \bibnamefont
      {Zubairy}},\ }\href@noop {} {\enquote {\bibinfo {title} {Quantum optics},}\ }
      (\bibinfo {year} {1999})\BibitemShut {NoStop}%
    \bibitem [{\citenamefont {Xiao}\ \emph {et~al.}(2020)\citenamefont {Xiao},
      \citenamefont {Deng}, \citenamefont {Wang}, \citenamefont {Zhu},
      \citenamefont {Wang}, \citenamefont {Yi},\ and\ \citenamefont
      {Xue}}]{xiao2020non}%
      \BibitemOpen
      \bibfield  {author} {\bibinfo {author} {\bibfnamefont {L.}~\bibnamefont
      {Xiao}}, \bibinfo {author} {\bibfnamefont {T.}~\bibnamefont {Deng}}, \bibinfo
      {author} {\bibfnamefont {K.}~\bibnamefont {Wang}}, \bibinfo {author}
      {\bibfnamefont {G.}~\bibnamefont {Zhu}}, \bibinfo {author} {\bibfnamefont
      {Z.}~\bibnamefont {Wang}}, \bibinfo {author} {\bibfnamefont {W.}~\bibnamefont
      {Yi}}, \ and\ \bibinfo {author} {\bibfnamefont {P.}~\bibnamefont {Xue}},\
      }\href@noop {} {\bibfield  {journal} {\bibinfo  {journal} {Nature Physics}\
      }\textbf {\bibinfo {volume} {16}},\ \bibinfo {pages} {761} (\bibinfo {year}
      {2020})}\BibitemShut {NoStop}%
    \bibitem [{\citenamefont {Zhang}\ \emph {et~al.}(2021)\citenamefont {Zhang},
      \citenamefont {Tian}, \citenamefont {Jiang}, \citenamefont {Lu},\ and\
      \citenamefont {Chen}}]{zhang2021observation}%
      \BibitemOpen
      \bibfield  {author} {\bibinfo {author} {\bibfnamefont {X.}~\bibnamefont
      {Zhang}}, \bibinfo {author} {\bibfnamefont {Y.}~\bibnamefont {Tian}},
      \bibinfo {author} {\bibfnamefont {J.-H.}\ \bibnamefont {Jiang}}, \bibinfo
      {author} {\bibfnamefont {M.-H.}\ \bibnamefont {Lu}}, \ and\ \bibinfo {author}
      {\bibfnamefont {Y.-F.}\ \bibnamefont {Chen}},\ }\href@noop {} {\bibfield
      {journal} {\bibinfo  {journal} {Nature communications}\ }\textbf {\bibinfo
      {volume} {12}},\ \bibinfo {pages} {1} (\bibinfo {year} {2021})}\BibitemShut
      {NoStop}%
    \bibitem [{\citenamefont {Zou}\ \emph {et~al.}(2021)\citenamefont {Zou},
      \citenamefont {Chen}, \citenamefont {He}, \citenamefont {Bao}, \citenamefont
      {Lee}, \citenamefont {Sun},\ and\ \citenamefont
      {Zhang}}]{zou2021observation}%
      \BibitemOpen
      \bibfield  {author} {\bibinfo {author} {\bibfnamefont {D.}~\bibnamefont
      {Zou}}, \bibinfo {author} {\bibfnamefont {T.}~\bibnamefont {Chen}}, \bibinfo
      {author} {\bibfnamefont {W.}~\bibnamefont {He}}, \bibinfo {author}
      {\bibfnamefont {J.}~\bibnamefont {Bao}}, \bibinfo {author} {\bibfnamefont
      {C.~H.}\ \bibnamefont {Lee}}, \bibinfo {author} {\bibfnamefont
      {H.}~\bibnamefont {Sun}}, \ and\ \bibinfo {author} {\bibfnamefont
      {X.}~\bibnamefont {Zhang}},\ }\href@noop {} {\bibfield  {journal} {\bibinfo
      {journal} {Nature Communications}\ }\textbf {\bibinfo {volume} {12}},\
      \bibinfo {pages} {1} (\bibinfo {year} {2021})}\BibitemShut {NoStop}%
    \bibitem [{\citenamefont {Bender}\ and\ \citenamefont
      {Boettcher}(1998)}]{bender1998real}%
      \BibitemOpen
      \bibfield  {author} {\bibinfo {author} {\bibfnamefont {C.~M.}\ \bibnamefont
      {Bender}}\ and\ \bibinfo {author} {\bibfnamefont {S.}~\bibnamefont
      {Boettcher}},\ }\href@noop {} {\bibfield  {journal} {\bibinfo  {journal}
      {Physical review letters}\ }\textbf {\bibinfo {volume} {80}},\ \bibinfo
      {pages} {5243} (\bibinfo {year} {1998})}\BibitemShut {NoStop}%
    \bibitem [{\citenamefont {Xia}\ \emph {et~al.}(2021)\citenamefont {Xia},
      \citenamefont {Kaltsas}, \citenamefont {Song}, \citenamefont {Komis},
      \citenamefont {Xu}, \citenamefont {Szameit}, \citenamefont {Buljan},
      \citenamefont {Makris},\ and\ \citenamefont {Chen}}]{xia2021nonlinear}%
      \BibitemOpen
      \bibfield  {author} {\bibinfo {author} {\bibfnamefont {S.}~\bibnamefont
      {Xia}}, \bibinfo {author} {\bibfnamefont {D.}~\bibnamefont {Kaltsas}},
      \bibinfo {author} {\bibfnamefont {D.}~\bibnamefont {Song}}, \bibinfo {author}
      {\bibfnamefont {I.}~\bibnamefont {Komis}}, \bibinfo {author} {\bibfnamefont
      {J.}~\bibnamefont {Xu}}, \bibinfo {author} {\bibfnamefont {A.}~\bibnamefont
      {Szameit}}, \bibinfo {author} {\bibfnamefont {H.}~\bibnamefont {Buljan}},
      \bibinfo {author} {\bibfnamefont {K.~G.}\ \bibnamefont {Makris}}, \ and\
      \bibinfo {author} {\bibfnamefont {Z.}~\bibnamefont {Chen}},\ }\href@noop {}
      {\bibfield  {journal} {\bibinfo  {journal} {Science}\ }\textbf {\bibinfo
      {volume} {372}},\ \bibinfo {pages} {72} (\bibinfo {year} {2021})}\BibitemShut
      {NoStop}%
    \bibitem [{\citenamefont {Feng}\ \emph {et~al.}(2017)\citenamefont {Feng},
      \citenamefont {El-Ganainy},\ and\ \citenamefont {Ge}}]{feng2017non}%
      \BibitemOpen
      \bibfield  {author} {\bibinfo {author} {\bibfnamefont {L.}~\bibnamefont
      {Feng}}, \bibinfo {author} {\bibfnamefont {R.}~\bibnamefont {El-Ganainy}}, \
      and\ \bibinfo {author} {\bibfnamefont {L.}~\bibnamefont {Ge}},\ }\href@noop
      {} {\bibfield  {journal} {\bibinfo  {journal} {Nature Photonics}\ }\textbf
      {\bibinfo {volume} {11}},\ \bibinfo {pages} {752} (\bibinfo {year}
      {2017})}\BibitemShut {NoStop}%
    \bibitem [{\citenamefont {Xiao}\ \emph {et~al.}(2017)\citenamefont {Xiao},
      \citenamefont {Zhan}, \citenamefont {Bian}, \citenamefont {Wang},
      \citenamefont {Zhang}, \citenamefont {Wang}, \citenamefont {Li},
      \citenamefont {Mochizuki}, \citenamefont {Kim}, \citenamefont {Kawakami}
      \emph {et~al.}}]{xiao2017observation}%
      \BibitemOpen
      \bibfield  {author} {\bibinfo {author} {\bibfnamefont {L.}~\bibnamefont
      {Xiao}}, \bibinfo {author} {\bibfnamefont {X.}~\bibnamefont {Zhan}}, \bibinfo
      {author} {\bibfnamefont {Z.}~\bibnamefont {Bian}}, \bibinfo {author}
      {\bibfnamefont {K.}~\bibnamefont {Wang}}, \bibinfo {author} {\bibfnamefont
      {X.}~\bibnamefont {Zhang}}, \bibinfo {author} {\bibfnamefont
      {X.}~\bibnamefont {Wang}}, \bibinfo {author} {\bibfnamefont {J.}~\bibnamefont
      {Li}}, \bibinfo {author} {\bibfnamefont {K.}~\bibnamefont {Mochizuki}},
      \bibinfo {author} {\bibfnamefont {D.}~\bibnamefont {Kim}}, \bibinfo {author}
      {\bibfnamefont {N.}~\bibnamefont {Kawakami}},  \emph {et~al.},\ }\href@noop
      {} {\bibfield  {journal} {\bibinfo  {journal} {Nature Physics}\ }\textbf
      {\bibinfo {volume} {13}},\ \bibinfo {pages} {1117} (\bibinfo {year}
      {2017})}\BibitemShut {NoStop}%
    \bibitem [{\citenamefont {Qi}\ and\ \citenamefont
      {Zhang}(2011)}]{qi2011topological}%
      \BibitemOpen
      \bibfield  {author} {\bibinfo {author} {\bibfnamefont {X.-L.}\ \bibnamefont
      {Qi}}\ and\ \bibinfo {author} {\bibfnamefont {S.-C.}\ \bibnamefont {Zhang}},\
      }\href@noop {} {\bibfield  {journal} {\bibinfo  {journal} {Reviews of Modern
      Physics}\ }\textbf {\bibinfo {volume} {83}},\ \bibinfo {pages} {1057}
      (\bibinfo {year} {2011})}\BibitemShut {NoStop}%
    \bibitem [{\citenamefont {Hasan}\ and\ \citenamefont
      {Kane}(2010)}]{hasan2010colloquium}%
      \BibitemOpen
      \bibfield  {author} {\bibinfo {author} {\bibfnamefont {M.~Z.}\ \bibnamefont
      {Hasan}}\ and\ \bibinfo {author} {\bibfnamefont {C.~L.}\ \bibnamefont
      {Kane}},\ }\href@noop {} {\bibfield  {journal} {\bibinfo  {journal} {Reviews
      of modern physics}\ }\textbf {\bibinfo {volume} {82}},\ \bibinfo {pages}
      {3045} (\bibinfo {year} {2010})}\BibitemShut {NoStop}%
    \bibitem [{\citenamefont {Asb{\'o}th}\ \emph {et~al.}(2016)\citenamefont
      {Asb{\'o}th}, \citenamefont {Oroszl{\'a}ny},\ and\ \citenamefont
      {P{\'a}lyi}}]{asboth2016short}%
      \BibitemOpen
      \bibfield  {author} {\bibinfo {author} {\bibfnamefont {J.~K.}\ \bibnamefont
      {Asb{\'o}th}}, \bibinfo {author} {\bibfnamefont {L.}~\bibnamefont
      {Oroszl{\'a}ny}}, \ and\ \bibinfo {author} {\bibfnamefont {A.}~\bibnamefont
      {P{\'a}lyi}},\ }\href@noop {} {\bibfield  {journal} {\bibinfo  {journal}
      {Lecture notes in physics}\ }\textbf {\bibinfo {volume} {919}},\ \bibinfo
      {pages} {166} (\bibinfo {year} {2016})}\BibitemShut {NoStop}%
    \bibitem [{\citenamefont {Shen}(2012)}]{shen2012topological}%
      \BibitemOpen
      \bibfield  {author} {\bibinfo {author} {\bibfnamefont {S.-Q.}\ \bibnamefont
      {Shen}},\ }\href@noop {} {\emph {\bibinfo {title} {Topological
      insulators}}},\ Vol.\ \bibinfo {volume} {174}\ (\bibinfo  {publisher}
      {Springer},\ \bibinfo {year} {2012})\BibitemShut {NoStop}%
    \bibitem [{\citenamefont {Okuma}\ \emph {et~al.}(2020)\citenamefont {Okuma},
      \citenamefont {Kawabata}, \citenamefont {Shiozaki},\ and\ \citenamefont
      {Sato}}]{okuma2020topological}%
      \BibitemOpen
      \bibfield  {author} {\bibinfo {author} {\bibfnamefont {N.}~\bibnamefont
      {Okuma}}, \bibinfo {author} {\bibfnamefont {K.}~\bibnamefont {Kawabata}},
      \bibinfo {author} {\bibfnamefont {K.}~\bibnamefont {Shiozaki}}, \ and\
      \bibinfo {author} {\bibfnamefont {M.}~\bibnamefont {Sato}},\ }\href@noop {}
      {\bibfield  {journal} {\bibinfo  {journal} {Physical review letters}\
      }\textbf {\bibinfo {volume} {124}},\ \bibinfo {pages} {086801} (\bibinfo
      {year} {2020})}\BibitemShut {NoStop}%
    \bibitem [{\citenamefont {Yao}\ and\ \citenamefont {Wang}(2018)}]{yao2018edge}%
      \BibitemOpen
      \bibfield  {author} {\bibinfo {author} {\bibfnamefont {S.}~\bibnamefont
      {Yao}}\ and\ \bibinfo {author} {\bibfnamefont {Z.}~\bibnamefont {Wang}},\
      }\href@noop {} {\bibfield  {journal} {\bibinfo  {journal} {Physical review
      letters}\ }\textbf {\bibinfo {volume} {121}},\ \bibinfo {pages} {086803}
      (\bibinfo {year} {2018})}\BibitemShut {NoStop}%
    \bibitem [{\citenamefont {Yu-Min}\ \emph {et~al.}(2021)\citenamefont {Yu-Min},
      \citenamefont {Fei},\ and\ \citenamefont {Zhong}}]{yu2021generalized}%
      \BibitemOpen
      \bibfield  {author} {\bibinfo {author} {\bibfnamefont {H.}~\bibnamefont
      {Yu-Min}}, \bibinfo {author} {\bibfnamefont {S.}~\bibnamefont {Fei}}, \ and\
      \bibinfo {author} {\bibfnamefont {W.}~\bibnamefont {Zhong}},\ }\href@noop {}
      {\bibfield  {journal} {\bibinfo  {journal} {ACTA PHYSICA SINICA}\ }\textbf
      {\bibinfo {volume} {70}} (\bibinfo {year} {2021})}\BibitemShut {NoStop}%
    \bibitem [{\citenamefont {Longhi}(2019)}]{longhi2019probing}%
      \BibitemOpen
      \bibfield  {author} {\bibinfo {author} {\bibfnamefont {S.}~\bibnamefont
      {Longhi}},\ }\href@noop {} {\bibfield  {journal} {\bibinfo  {journal}
      {Physical Review Research}\ }\textbf {\bibinfo {volume} {1}},\ \bibinfo
      {pages} {023013} (\bibinfo {year} {2019})}\BibitemShut {NoStop}%
    \bibitem [{\citenamefont {Li}\ \emph {et~al.}(2020)\citenamefont {Li},
      \citenamefont {Lee}, \citenamefont {Mu},\ and\ \citenamefont
      {Gong}}]{li2020critical}%
      \BibitemOpen
      \bibfield  {author} {\bibinfo {author} {\bibfnamefont {L.}~\bibnamefont
      {Li}}, \bibinfo {author} {\bibfnamefont {C.~H.}\ \bibnamefont {Lee}},
      \bibinfo {author} {\bibfnamefont {S.}~\bibnamefont {Mu}}, \ and\ \bibinfo
      {author} {\bibfnamefont {J.}~\bibnamefont {Gong}},\ }\href@noop {} {\bibfield
       {journal} {\bibinfo  {journal} {Nature communications}\ }\textbf {\bibinfo
      {volume} {11}},\ \bibinfo {pages} {1} (\bibinfo {year} {2020})}\BibitemShut
      {NoStop}%
    \bibitem [{\citenamefont {Yokomizo}\ and\ \citenamefont
      {Murakami}(2019)}]{yokomizo2019non}%
      \BibitemOpen
      \bibfield  {author} {\bibinfo {author} {\bibfnamefont {K.}~\bibnamefont
      {Yokomizo}}\ and\ \bibinfo {author} {\bibfnamefont {S.}~\bibnamefont
      {Murakami}},\ }\href@noop {} {\bibfield  {journal} {\bibinfo  {journal}
      {Physical review letters}\ }\textbf {\bibinfo {volume} {123}},\ \bibinfo
      {pages} {066404} (\bibinfo {year} {2019})}\BibitemShut {NoStop}%
    \bibitem [{\citenamefont {Kawabata}\ \emph {et~al.}(2020)\citenamefont
      {Kawabata}, \citenamefont {Okuma},\ and\ \citenamefont
      {Sato}}]{kawabata2020non}%
      \BibitemOpen
      \bibfield  {author} {\bibinfo {author} {\bibfnamefont {K.}~\bibnamefont
      {Kawabata}}, \bibinfo {author} {\bibfnamefont {N.}~\bibnamefont {Okuma}}, \
      and\ \bibinfo {author} {\bibfnamefont {M.}~\bibnamefont {Sato}},\ }\href@noop
      {} {\bibfield  {journal} {\bibinfo  {journal} {Physical Review B}\ }\textbf
      {\bibinfo {volume} {101}},\ \bibinfo {pages} {195147} (\bibinfo {year}
      {2020})}\BibitemShut {NoStop}%
    \bibitem [{\citenamefont {Yao}\ \emph {et~al.}(2018)\citenamefont {Yao},
      \citenamefont {Song},\ and\ \citenamefont {Wang}}]{yao2018non}%
      \BibitemOpen
      \bibfield  {author} {\bibinfo {author} {\bibfnamefont {S.}~\bibnamefont
      {Yao}}, \bibinfo {author} {\bibfnamefont {F.}~\bibnamefont {Song}}, \ and\
      \bibinfo {author} {\bibfnamefont {Z.}~\bibnamefont {Wang}},\ }\href@noop {}
      {\bibfield  {journal} {\bibinfo  {journal} {Physical review letters}\
      }\textbf {\bibinfo {volume} {121}},\ \bibinfo {pages} {136802} (\bibinfo
      {year} {2018})}\BibitemShut {NoStop}%
    \bibitem [{\citenamefont {Yang}\ \emph {et~al.}(2020)\citenamefont {Yang},
      \citenamefont {Zhang}, \citenamefont {Fang},\ and\ \citenamefont
      {Hu}}]{yang2020non}%
      \BibitemOpen
      \bibfield  {author} {\bibinfo {author} {\bibfnamefont {Z.}~\bibnamefont
      {Yang}}, \bibinfo {author} {\bibfnamefont {K.}~\bibnamefont {Zhang}},
      \bibinfo {author} {\bibfnamefont {C.}~\bibnamefont {Fang}}, \ and\ \bibinfo
      {author} {\bibfnamefont {J.}~\bibnamefont {Hu}},\ }\href@noop {} {\bibfield
      {journal} {\bibinfo  {journal} {Physical Review Letters}\ }\textbf {\bibinfo
      {volume} {125}},\ \bibinfo {pages} {226402} (\bibinfo {year}
      {2020})}\BibitemShut {NoStop}%
    \bibitem [{\citenamefont {Zhang}\ \emph {et~al.}(2020)\citenamefont {Zhang},
      \citenamefont {Yang},\ and\ \citenamefont {Fang}}]{zhang2020correspondence}%
      \BibitemOpen
      \bibfield  {author} {\bibinfo {author} {\bibfnamefont {K.}~\bibnamefont
      {Zhang}}, \bibinfo {author} {\bibfnamefont {Z.}~\bibnamefont {Yang}}, \ and\
      \bibinfo {author} {\bibfnamefont {C.}~\bibnamefont {Fang}},\ }\href@noop {}
      {\bibfield  {journal} {\bibinfo  {journal} {Physical Review Letters}\
      }\textbf {\bibinfo {volume} {125}},\ \bibinfo {pages} {126402} (\bibinfo
      {year} {2020})}\BibitemShut {NoStop}%
    \bibitem [{\citenamefont {Longhi}(2020)}]{longhi2020non}%
      \BibitemOpen
      \bibfield  {author} {\bibinfo {author} {\bibfnamefont {S.}~\bibnamefont
      {Longhi}},\ }\href@noop {} {\bibfield  {journal} {\bibinfo  {journal}
      {Physical Review Letters}\ }\textbf {\bibinfo {volume} {124}},\ \bibinfo
      {pages} {066602} (\bibinfo {year} {2020})}\BibitemShut {NoStop}%
    \bibitem [{\citenamefont {Xue}\ \emph {et~al.}(2021)\citenamefont {Xue},
      \citenamefont {Li}, \citenamefont {Hu}, \citenamefont {Song},\ and\
      \citenamefont {Wang}}]{xue2021simple}%
      \BibitemOpen
      \bibfield  {author} {\bibinfo {author} {\bibfnamefont {W.-T.}\ \bibnamefont
      {Xue}}, \bibinfo {author} {\bibfnamefont {M.-R.}\ \bibnamefont {Li}},
      \bibinfo {author} {\bibfnamefont {Y.-M.}\ \bibnamefont {Hu}}, \bibinfo
      {author} {\bibfnamefont {F.}~\bibnamefont {Song}}, \ and\ \bibinfo {author}
      {\bibfnamefont {Z.}~\bibnamefont {Wang}},\ }\href@noop {} {\bibfield
      {journal} {\bibinfo  {journal} {Physical Review B}\ }\textbf {\bibinfo
      {volume} {103}},\ \bibinfo {pages} {L241408} (\bibinfo {year}
      {2021})}\BibitemShut {NoStop}%
    \bibitem [{\citenamefont {Feng}\ \emph {et~al.}(2011)\citenamefont {Feng},
      \citenamefont {Ayache}, \citenamefont {Huang}, \citenamefont {Xu},
      \citenamefont {Lu}, \citenamefont {Chen}, \citenamefont {Fainman},\ and\
      \citenamefont {Scherer}}]{feng2011nonreciprocal}%
      \BibitemOpen
      \bibfield  {author} {\bibinfo {author} {\bibfnamefont {L.}~\bibnamefont
      {Feng}}, \bibinfo {author} {\bibfnamefont {M.}~\bibnamefont {Ayache}},
      \bibinfo {author} {\bibfnamefont {J.}~\bibnamefont {Huang}}, \bibinfo
      {author} {\bibfnamefont {Y.-L.}\ \bibnamefont {Xu}}, \bibinfo {author}
      {\bibfnamefont {M.-H.}\ \bibnamefont {Lu}}, \bibinfo {author} {\bibfnamefont
      {Y.-F.}\ \bibnamefont {Chen}}, \bibinfo {author} {\bibfnamefont
      {Y.}~\bibnamefont {Fainman}}, \ and\ \bibinfo {author} {\bibfnamefont
      {A.}~\bibnamefont {Scherer}},\ }\href@noop {} {\bibfield  {journal} {\bibinfo
       {journal} {Science}\ }\textbf {\bibinfo {volume} {333}},\ \bibinfo {pages}
      {729} (\bibinfo {year} {2011})}\BibitemShut {NoStop}%
    \bibitem [{\citenamefont {Bi}\ \emph {et~al.}(2011)\citenamefont {Bi},
      \citenamefont {Hu}, \citenamefont {Jiang}, \citenamefont {Kim}, \citenamefont
      {Dionne}, \citenamefont {Kimerling},\ and\ \citenamefont
      {Ross}}]{bi2011chip}%
      \BibitemOpen
      \bibfield  {author} {\bibinfo {author} {\bibfnamefont {L.}~\bibnamefont
      {Bi}}, \bibinfo {author} {\bibfnamefont {J.}~\bibnamefont {Hu}}, \bibinfo
      {author} {\bibfnamefont {P.}~\bibnamefont {Jiang}}, \bibinfo {author}
      {\bibfnamefont {D.~H.}\ \bibnamefont {Kim}}, \bibinfo {author} {\bibfnamefont
      {G.~F.}\ \bibnamefont {Dionne}}, \bibinfo {author} {\bibfnamefont {L.~C.}\
      \bibnamefont {Kimerling}}, \ and\ \bibinfo {author} {\bibfnamefont
      {C.}~\bibnamefont {Ross}},\ }\href@noop {} {\bibfield  {journal} {\bibinfo
      {journal} {Nature Photonics}\ }\textbf {\bibinfo {volume} {5}},\ \bibinfo
      {pages} {758} (\bibinfo {year} {2011})}\BibitemShut {NoStop}%
    \bibitem [{\citenamefont {Ozawa}\ \emph {et~al.}(2019)\citenamefont {Ozawa},
      \citenamefont {Price}, \citenamefont {Amo}, \citenamefont {Goldman},
      \citenamefont {Hafezi}, \citenamefont {Lu}, \citenamefont {Rechtsman},
      \citenamefont {Schuster}, \citenamefont {Simon}, \citenamefont {Zilberberg}
      \emph {et~al.}}]{ozawa2019topological}%
      \BibitemOpen
      \bibfield  {author} {\bibinfo {author} {\bibfnamefont {T.}~\bibnamefont
      {Ozawa}}, \bibinfo {author} {\bibfnamefont {H.~M.}\ \bibnamefont {Price}},
      \bibinfo {author} {\bibfnamefont {A.}~\bibnamefont {Amo}}, \bibinfo {author}
      {\bibfnamefont {N.}~\bibnamefont {Goldman}}, \bibinfo {author} {\bibfnamefont
      {M.}~\bibnamefont {Hafezi}}, \bibinfo {author} {\bibfnamefont
      {L.}~\bibnamefont {Lu}}, \bibinfo {author} {\bibfnamefont {M.~C.}\
      \bibnamefont {Rechtsman}}, \bibinfo {author} {\bibfnamefont {D.}~\bibnamefont
      {Schuster}}, \bibinfo {author} {\bibfnamefont {J.}~\bibnamefont {Simon}},
      \bibinfo {author} {\bibfnamefont {O.}~\bibnamefont {Zilberberg}},  \emph
      {et~al.},\ }\href@noop {} {\bibfield  {journal} {\bibinfo  {journal} {Reviews
      of Modern Physics}\ }\textbf {\bibinfo {volume} {91}},\ \bibinfo {pages}
      {015006} (\bibinfo {year} {2019})}\BibitemShut {NoStop}%
    \bibitem [{\citenamefont {Lu}\ \emph {et~al.}(2014)\citenamefont {Lu},
      \citenamefont {Joannopoulos},\ and\ \citenamefont
      {Solja{\v{c}}i{\'c}}}]{lu2014topological}%
      \BibitemOpen
      \bibfield  {author} {\bibinfo {author} {\bibfnamefont {L.}~\bibnamefont
      {Lu}}, \bibinfo {author} {\bibfnamefont {J.~D.}\ \bibnamefont
      {Joannopoulos}}, \ and\ \bibinfo {author} {\bibfnamefont {M.}~\bibnamefont
      {Solja{\v{c}}i{\'c}}},\ }\href@noop {} {\bibfield  {journal} {\bibinfo
      {journal} {Nature photonics}\ }\textbf {\bibinfo {volume} {8}},\ \bibinfo
      {pages} {821} (\bibinfo {year} {2014})}\BibitemShut {NoStop}%
    \bibitem [{\citenamefont {Smirnova}\ \emph {et~al.}(2020)\citenamefont
      {Smirnova}, \citenamefont {Leykam}, \citenamefont {Chong},\ and\
      \citenamefont {Kivshar}}]{smirnova2020nonlinear}%
      \BibitemOpen
      \bibfield  {author} {\bibinfo {author} {\bibfnamefont {D.}~\bibnamefont
      {Smirnova}}, \bibinfo {author} {\bibfnamefont {D.}~\bibnamefont {Leykam}},
      \bibinfo {author} {\bibfnamefont {Y.}~\bibnamefont {Chong}}, \ and\ \bibinfo
      {author} {\bibfnamefont {Y.}~\bibnamefont {Kivshar}},\ }\href@noop {}
      {\bibfield  {journal} {\bibinfo  {journal} {Applied Physics Reviews}\
      }\textbf {\bibinfo {volume} {7}},\ \bibinfo {pages} {021306} (\bibinfo {year}
      {2020})}\BibitemShut {NoStop}%
    \bibitem [{\citenamefont {Khanikaev}\ and\ \citenamefont
      {Shvets}(2017)}]{khanikaev2017two}%
      \BibitemOpen
      \bibfield  {author} {\bibinfo {author} {\bibfnamefont {A.~B.}\ \bibnamefont
      {Khanikaev}}\ and\ \bibinfo {author} {\bibfnamefont {G.}~\bibnamefont
      {Shvets}},\ }\href@noop {} {\bibfield  {journal} {\bibinfo  {journal} {Nature
      photonics}\ }\textbf {\bibinfo {volume} {11}},\ \bibinfo {pages} {763}
      (\bibinfo {year} {2017})}\BibitemShut {NoStop}%
    \bibitem [{\citenamefont {Kim}\ \emph {et~al.}(2020)\citenamefont {Kim},
      \citenamefont {Jacob},\ and\ \citenamefont {Rho}}]{kim2020recent}%
      \BibitemOpen
      \bibfield  {author} {\bibinfo {author} {\bibfnamefont {M.}~\bibnamefont
      {Kim}}, \bibinfo {author} {\bibfnamefont {Z.}~\bibnamefont {Jacob}}, \ and\
      \bibinfo {author} {\bibfnamefont {J.}~\bibnamefont {Rho}},\ }\href@noop {}
      {\bibfield  {journal} {\bibinfo  {journal} {Light: Science \& Applications}\
      }\textbf {\bibinfo {volume} {9}},\ \bibinfo {pages} {1} (\bibinfo {year}
      {2020})}\BibitemShut {NoStop}%
    \bibitem [{\citenamefont {Aspelmeyer}\ \emph {et~al.}(2014)\citenamefont
      {Aspelmeyer}, \citenamefont {Kippenberg},\ and\ \citenamefont
      {Marquardt}}]{aspelmeyer2014cavity}%
      \BibitemOpen
      \bibfield  {author} {\bibinfo {author} {\bibfnamefont {M.}~\bibnamefont
      {Aspelmeyer}}, \bibinfo {author} {\bibfnamefont {T.~J.}\ \bibnamefont
      {Kippenberg}}, \ and\ \bibinfo {author} {\bibfnamefont {F.}~\bibnamefont
      {Marquardt}},\ }\href@noop {} {\bibfield  {journal} {\bibinfo  {journal}
      {Reviews of Modern Physics}\ }\textbf {\bibinfo {volume} {86}},\ \bibinfo
      {pages} {1391} (\bibinfo {year} {2014})}\BibitemShut {NoStop}%
    \bibitem [{\citenamefont {Marquardt}\ and\ \citenamefont
      {Girvin}(2009)}]{marquardt2009optomechanics}%
      \BibitemOpen
      \bibfield  {author} {\bibinfo {author} {\bibfnamefont {F.}~\bibnamefont
      {Marquardt}}\ and\ \bibinfo {author} {\bibfnamefont {S.~M.}\ \bibnamefont
      {Girvin}},\ }\href@noop {} {\bibfield  {journal} {\bibinfo  {journal}
      {Physics}\ }\textbf {\bibinfo {volume} {2}},\ \bibinfo {pages} {40} (\bibinfo
      {year} {2009})}\BibitemShut {NoStop}%
    \bibitem [{\citenamefont {Kippenberg}\ and\ \citenamefont
      {Vahala}(2008)}]{kippenberg2008cavity}%
      \BibitemOpen
      \bibfield  {author} {\bibinfo {author} {\bibfnamefont {T.~J.}\ \bibnamefont
      {Kippenberg}}\ and\ \bibinfo {author} {\bibfnamefont {K.~J.}\ \bibnamefont
      {Vahala}},\ }\href@noop {} {\bibfield  {journal} {\bibinfo  {journal}
      {science}\ }\textbf {\bibinfo {volume} {321}},\ \bibinfo {pages} {1172}
      (\bibinfo {year} {2008})}\BibitemShut {NoStop}%
    \bibitem [{\citenamefont {Metcalfe}(2014)}]{metcalfe2014applications}%
      \BibitemOpen
      \bibfield  {author} {\bibinfo {author} {\bibfnamefont {M.}~\bibnamefont
      {Metcalfe}},\ }\href@noop {} {\bibfield  {journal} {\bibinfo  {journal}
      {Applied Physics Reviews}\ }\textbf {\bibinfo {volume} {1}},\ \bibinfo
      {pages} {031105} (\bibinfo {year} {2014})}\BibitemShut {NoStop}%
    \bibitem [{\citenamefont {Kippenberg}\ and\ \citenamefont
      {Vahala}(2007)}]{kippenberg2007cavity}%
      \BibitemOpen
      \bibfield  {author} {\bibinfo {author} {\bibfnamefont {T.~J.}\ \bibnamefont
      {Kippenberg}}\ and\ \bibinfo {author} {\bibfnamefont {K.~J.}\ \bibnamefont
      {Vahala}},\ }\href@noop {} {\bibfield  {journal} {\bibinfo  {journal} {Optics
      express}\ }\textbf {\bibinfo {volume} {15}},\ \bibinfo {pages} {17172}
      (\bibinfo {year} {2007})}\BibitemShut {NoStop}%
    \bibitem [{\citenamefont {Dong}\ \emph {et~al.}(2012)\citenamefont {Dong},
      \citenamefont {Fiore}, \citenamefont {Kuzyk},\ and\ \citenamefont
      {Wang}}]{dong2012optomechanical}%
      \BibitemOpen
      \bibfield  {author} {\bibinfo {author} {\bibfnamefont {C.}~\bibnamefont
      {Dong}}, \bibinfo {author} {\bibfnamefont {V.}~\bibnamefont {Fiore}},
      \bibinfo {author} {\bibfnamefont {M.~C.}\ \bibnamefont {Kuzyk}}, \ and\
      \bibinfo {author} {\bibfnamefont {H.}~\bibnamefont {Wang}},\ }\href@noop {}
      {\bibfield  {journal} {\bibinfo  {journal} {Science}\ }\textbf {\bibinfo
      {volume} {338}},\ \bibinfo {pages} {1609} (\bibinfo {year}
      {2012})}\BibitemShut {NoStop}%
    \bibitem [{\citenamefont {Marangos}(1998)}]{marangos1998electromagnetically}%
      \BibitemOpen
      \bibfield  {author} {\bibinfo {author} {\bibfnamefont {J.~P.}\ \bibnamefont
      {Marangos}},\ }\href@noop {} {\bibfield  {journal} {\bibinfo  {journal}
      {Journal of modern optics}\ }\textbf {\bibinfo {volume} {45}},\ \bibinfo
      {pages} {471} (\bibinfo {year} {1998})}\BibitemShut {NoStop}%
    \bibitem [{\citenamefont {Weis}\ \emph {et~al.}(2010)\citenamefont {Weis},
      \citenamefont {Rivi{\`e}re}, \citenamefont {Del{\'e}glise}, \citenamefont
      {Gavartin}, \citenamefont {Arcizet}, \citenamefont {Schliesser},\ and\
      \citenamefont {Kippenberg}}]{weis2010optomechanically}%
      \BibitemOpen
      \bibfield  {author} {\bibinfo {author} {\bibfnamefont {S.}~\bibnamefont
      {Weis}}, \bibinfo {author} {\bibfnamefont {R.}~\bibnamefont {Rivi{\`e}re}},
      \bibinfo {author} {\bibfnamefont {S.}~\bibnamefont {Del{\'e}glise}}, \bibinfo
      {author} {\bibfnamefont {E.}~\bibnamefont {Gavartin}}, \bibinfo {author}
      {\bibfnamefont {O.}~\bibnamefont {Arcizet}}, \bibinfo {author} {\bibfnamefont
      {A.}~\bibnamefont {Schliesser}}, \ and\ \bibinfo {author} {\bibfnamefont
      {T.~J.}\ \bibnamefont {Kippenberg}},\ }\href@noop {} {\bibfield  {journal}
      {\bibinfo  {journal} {Science}\ }\textbf {\bibinfo {volume} {330}},\ \bibinfo
      {pages} {1520} (\bibinfo {year} {2010})}\BibitemShut {NoStop}%
    \bibitem [{\citenamefont {Xiong}\ and\ \citenamefont
      {Wu}(2018)}]{xiong2018fundamentals}%
      \BibitemOpen
      \bibfield  {author} {\bibinfo {author} {\bibfnamefont {H.}~\bibnamefont
      {Xiong}}\ and\ \bibinfo {author} {\bibfnamefont {Y.}~\bibnamefont {Wu}},\
      }\href@noop {} {\bibfield  {journal} {\bibinfo  {journal} {Applied Physics
      Reviews}\ }\textbf {\bibinfo {volume} {5}},\ \bibinfo {pages} {031305}
      (\bibinfo {year} {2018})}\BibitemShut {NoStop}%
    \bibitem [{\citenamefont {Kronwald}\ and\ \citenamefont
      {Marquardt}(2013)}]{kronwald2013optomechanically}%
      \BibitemOpen
      \bibfield  {author} {\bibinfo {author} {\bibfnamefont {A.}~\bibnamefont
      {Kronwald}}\ and\ \bibinfo {author} {\bibfnamefont {F.}~\bibnamefont
      {Marquardt}},\ }\href@noop {} {\bibfield  {journal} {\bibinfo  {journal}
      {Physical review letters}\ }\textbf {\bibinfo {volume} {111}},\ \bibinfo
      {pages} {133601} (\bibinfo {year} {2013})}\BibitemShut {NoStop}%
    \bibitem [{\citenamefont {Dong}\ \emph {et~al.}(2013)\citenamefont {Dong},
      \citenamefont {Fiore}, \citenamefont {Kuzyk},\ and\ \citenamefont
      {Wang}}]{dong2013transient}%
      \BibitemOpen
      \bibfield  {author} {\bibinfo {author} {\bibfnamefont {C.}~\bibnamefont
      {Dong}}, \bibinfo {author} {\bibfnamefont {V.}~\bibnamefont {Fiore}},
      \bibinfo {author} {\bibfnamefont {M.~C.}\ \bibnamefont {Kuzyk}}, \ and\
      \bibinfo {author} {\bibfnamefont {H.}~\bibnamefont {Wang}},\ }\href@noop {}
      {\bibfield  {journal} {\bibinfo  {journal} {Physical Review A}\ }\textbf
      {\bibinfo {volume} {87}},\ \bibinfo {pages} {055802} (\bibinfo {year}
      {2013})}\BibitemShut {NoStop}%
    \bibitem [{\citenamefont {L{\"u}}\ \emph {et~al.}(2018)\citenamefont {L{\"u}},
      \citenamefont {Wang}, \citenamefont {Yang},\ and\ \citenamefont
      {Jing}}]{lu2018optomechanically}%
      \BibitemOpen
      \bibfield  {author} {\bibinfo {author} {\bibfnamefont {H.}~\bibnamefont
      {L{\"u}}}, \bibinfo {author} {\bibfnamefont {C.}~\bibnamefont {Wang}},
      \bibinfo {author} {\bibfnamefont {L.}~\bibnamefont {Yang}}, \ and\ \bibinfo
      {author} {\bibfnamefont {H.}~\bibnamefont {Jing}},\ }\href@noop {} {\bibfield
       {journal} {\bibinfo  {journal} {Physical Review Applied}\ }\textbf {\bibinfo
      {volume} {10}},\ \bibinfo {pages} {014006} (\bibinfo {year}
      {2018})}\BibitemShut {NoStop}%
    \bibitem [{\citenamefont {Lai}\ \emph {et~al.}(2020)\citenamefont {Lai},
      \citenamefont {Wang}, \citenamefont {Qin}, \citenamefont {Hou}, \citenamefont
      {Nori},\ and\ \citenamefont {Liao}}]{lai2020tunable}%
      \BibitemOpen
      \bibfield  {author} {\bibinfo {author} {\bibfnamefont {D.-G.}\ \bibnamefont
      {Lai}}, \bibinfo {author} {\bibfnamefont {X.}~\bibnamefont {Wang}}, \bibinfo
      {author} {\bibfnamefont {W.}~\bibnamefont {Qin}}, \bibinfo {author}
      {\bibfnamefont {B.-P.}\ \bibnamefont {Hou}}, \bibinfo {author} {\bibfnamefont
      {F.}~\bibnamefont {Nori}}, \ and\ \bibinfo {author} {\bibfnamefont {J.-Q.}\
      \bibnamefont {Liao}},\ }\href@noop {} {\bibfield  {journal} {\bibinfo
      {journal} {Physical Review A}\ }\textbf {\bibinfo {volume} {102}},\ \bibinfo
      {pages} {023707} (\bibinfo {year} {2020})}\BibitemShut {NoStop}%
    \bibitem [{\citenamefont {L{\"u}}\ \emph {et~al.}(2017)\citenamefont {L{\"u}},
      \citenamefont {Jiang}, \citenamefont {Wang},\ and\ \citenamefont
      {Jing}}]{lu2017optomechanically}%
      \BibitemOpen
      \bibfield  {author} {\bibinfo {author} {\bibfnamefont {H.}~\bibnamefont
      {L{\"u}}}, \bibinfo {author} {\bibfnamefont {Y.}~\bibnamefont {Jiang}},
      \bibinfo {author} {\bibfnamefont {Y.-Z.}\ \bibnamefont {Wang}}, \ and\
      \bibinfo {author} {\bibfnamefont {H.}~\bibnamefont {Jing}},\ }\href@noop {}
      {\bibfield  {journal} {\bibinfo  {journal} {Photonics Research}\ }\textbf
      {\bibinfo {volume} {5}},\ \bibinfo {pages} {367} (\bibinfo {year}
      {2017})}\BibitemShut {NoStop}%
    \bibitem [{\citenamefont {Lei}\ \emph {et~al.}(2015)\citenamefont {Lei},
      \citenamefont {Gao}, \citenamefont {Du}, \citenamefont {Jing},\ and\
      \citenamefont {Long}}]{lei2015three}%
      \BibitemOpen
      \bibfield  {author} {\bibinfo {author} {\bibfnamefont {F.-C.}\ \bibnamefont
      {Lei}}, \bibinfo {author} {\bibfnamefont {M.}~\bibnamefont {Gao}}, \bibinfo
      {author} {\bibfnamefont {C.}~\bibnamefont {Du}}, \bibinfo {author}
      {\bibfnamefont {Q.-L.}\ \bibnamefont {Jing}}, \ and\ \bibinfo {author}
      {\bibfnamefont {G.-L.}\ \bibnamefont {Long}},\ }\href@noop {} {\bibfield
      {journal} {\bibinfo  {journal} {Optics express}\ }\textbf {\bibinfo {volume}
      {23}},\ \bibinfo {pages} {11508} (\bibinfo {year} {2015})}\BibitemShut
      {NoStop}%
    \bibitem [{\citenamefont {Qin}\ \emph {et~al.}(2020)\citenamefont {Qin},
      \citenamefont {Yang}, \citenamefont {Mao}, \citenamefont {Wen}, \citenamefont
      {Wang}, \citenamefont {Ruan},\ and\ \citenamefont
      {Long}}]{qin2020manipulation}%
      \BibitemOpen
      \bibfield  {author} {\bibinfo {author} {\bibfnamefont {G.-q.}\ \bibnamefont
      {Qin}}, \bibinfo {author} {\bibfnamefont {H.}~\bibnamefont {Yang}}, \bibinfo
      {author} {\bibfnamefont {X.}~\bibnamefont {Mao}}, \bibinfo {author}
      {\bibfnamefont {J.-w.}\ \bibnamefont {Wen}}, \bibinfo {author} {\bibfnamefont
      {M.}~\bibnamefont {Wang}}, \bibinfo {author} {\bibfnamefont {D.}~\bibnamefont
      {Ruan}}, \ and\ \bibinfo {author} {\bibfnamefont {G.-l.}\ \bibnamefont
      {Long}},\ }\href@noop {} {\bibfield  {journal} {\bibinfo  {journal} {Optics
      express}\ }\textbf {\bibinfo {volume} {28}},\ \bibinfo {pages} {580}
      (\bibinfo {year} {2020})}\BibitemShut {NoStop}%
    \bibitem [{\citenamefont {Mao}\ \emph {et~al.}(2022)\citenamefont {Mao},
      \citenamefont {Qin}, \citenamefont {Yang}, \citenamefont {Wang},
      \citenamefont {Wang}, \citenamefont {Li}, \citenamefont {Xue},\ and\
      \citenamefont {Long}}]{mao2022tunable}%
      \BibitemOpen
      \bibfield  {author} {\bibinfo {author} {\bibfnamefont {X.}~\bibnamefont
      {Mao}}, \bibinfo {author} {\bibfnamefont {G.-Q.}\ \bibnamefont {Qin}},
      \bibinfo {author} {\bibfnamefont {H.}~\bibnamefont {Yang}}, \bibinfo {author}
      {\bibfnamefont {Z.}~\bibnamefont {Wang}}, \bibinfo {author} {\bibfnamefont
      {M.}~\bibnamefont {Wang}}, \bibinfo {author} {\bibfnamefont {G.-Q.}\
      \bibnamefont {Li}}, \bibinfo {author} {\bibfnamefont {P.}~\bibnamefont
      {Xue}}, \ and\ \bibinfo {author} {\bibfnamefont {G.-L.}\ \bibnamefont
      {Long}},\ }\href@noop {} {\bibfield  {journal} {\bibinfo  {journal} {Physical
      Review A}\ }\textbf {\bibinfo {volume} {105}},\ \bibinfo {pages} {033526}
      (\bibinfo {year} {2022})}\BibitemShut {NoStop}%
    \bibitem [{\citenamefont {Peng}\ \emph {et~al.}(2014)\citenamefont {Peng},
      \citenamefont {{\"O}zdemir}, \citenamefont {Lei}, \citenamefont {Monifi},
      \citenamefont {Gianfreda}, \citenamefont {Long}, \citenamefont {Fan},
      \citenamefont {Nori}, \citenamefont {Bender},\ and\ \citenamefont
      {Yang}}]{peng2014parity}%
      \BibitemOpen
      \bibfield  {author} {\bibinfo {author} {\bibfnamefont {B.}~\bibnamefont
      {Peng}}, \bibinfo {author} {\bibfnamefont {{\c{S}}.~K.}\ \bibnamefont
      {{\"O}zdemir}}, \bibinfo {author} {\bibfnamefont {F.}~\bibnamefont {Lei}},
      \bibinfo {author} {\bibfnamefont {F.}~\bibnamefont {Monifi}}, \bibinfo
      {author} {\bibfnamefont {M.}~\bibnamefont {Gianfreda}}, \bibinfo {author}
      {\bibfnamefont {G.~L.}\ \bibnamefont {Long}}, \bibinfo {author}
      {\bibfnamefont {S.}~\bibnamefont {Fan}}, \bibinfo {author} {\bibfnamefont
      {F.}~\bibnamefont {Nori}}, \bibinfo {author} {\bibfnamefont {C.~M.}\
      \bibnamefont {Bender}}, \ and\ \bibinfo {author} {\bibfnamefont
      {L.}~\bibnamefont {Yang}},\ }\href@noop {} {\bibfield  {journal} {\bibinfo
      {journal} {Nature Physics}\ }\textbf {\bibinfo {volume} {10}},\ \bibinfo
      {pages} {394} (\bibinfo {year} {2014})}\BibitemShut {NoStop}%
    \bibitem [{\citenamefont {Xie}\ \emph {et~al.}(2021)\citenamefont {Xie},
      \citenamefont {Qin}, \citenamefont {Zhang}, \citenamefont {Wang},
      \citenamefont {Li}, \citenamefont {Ruan},\ and\ \citenamefont
      {Long}}]{xie2021phase}%
      \BibitemOpen
      \bibfield  {author} {\bibinfo {author} {\bibfnamefont {R.-R.}\ \bibnamefont
      {Xie}}, \bibinfo {author} {\bibfnamefont {G.-Q.}\ \bibnamefont {Qin}},
      \bibinfo {author} {\bibfnamefont {H.}~\bibnamefont {Zhang}}, \bibinfo
      {author} {\bibfnamefont {M.}~\bibnamefont {Wang}}, \bibinfo {author}
      {\bibfnamefont {G.-Q.}\ \bibnamefont {Li}}, \bibinfo {author} {\bibfnamefont
      {D.}~\bibnamefont {Ruan}}, \ and\ \bibinfo {author} {\bibfnamefont {G.-L.}\
      \bibnamefont {Long}},\ }\href@noop {} {\bibfield  {journal} {\bibinfo
      {journal} {Optics Letters}\ }\textbf {\bibinfo {volume} {46}},\ \bibinfo
      {pages} {773} (\bibinfo {year} {2021})}\BibitemShut {NoStop}%
    \bibitem [{\citenamefont {Jiang}\ \emph {et~al.}(2015)\citenamefont {Jiang},
      \citenamefont {Wang}, \citenamefont {Kuzyk}, \citenamefont {Oo},
      \citenamefont {Long},\ and\ \citenamefont {Wang}}]{jiang2015chip}%
      \BibitemOpen
      \bibfield  {author} {\bibinfo {author} {\bibfnamefont {X.}~\bibnamefont
      {Jiang}}, \bibinfo {author} {\bibfnamefont {M.}~\bibnamefont {Wang}},
      \bibinfo {author} {\bibfnamefont {M.~C.}\ \bibnamefont {Kuzyk}}, \bibinfo
      {author} {\bibfnamefont {T.}~\bibnamefont {Oo}}, \bibinfo {author}
      {\bibfnamefont {G.-L.}\ \bibnamefont {Long}}, \ and\ \bibinfo {author}
      {\bibfnamefont {H.}~\bibnamefont {Wang}},\ }\href@noop {} {\bibfield
      {journal} {\bibinfo  {journal} {Optics express}\ }\textbf {\bibinfo {volume}
      {23}},\ \bibinfo {pages} {27260} (\bibinfo {year} {2015})}\BibitemShut
      {NoStop}%
    \bibitem [{\citenamefont {Mao}\ \emph {et~al.}(2020)\citenamefont {Mao},
      \citenamefont {Qin}, \citenamefont {Yang}, \citenamefont {Zhang},
      \citenamefont {Wang},\ and\ \citenamefont {Long}}]{mao2020enhanced}%
      \BibitemOpen
      \bibfield  {author} {\bibinfo {author} {\bibfnamefont {X.}~\bibnamefont
      {Mao}}, \bibinfo {author} {\bibfnamefont {G.-Q.}\ \bibnamefont {Qin}},
      \bibinfo {author} {\bibfnamefont {H.}~\bibnamefont {Yang}}, \bibinfo {author}
      {\bibfnamefont {H.}~\bibnamefont {Zhang}}, \bibinfo {author} {\bibfnamefont
      {M.}~\bibnamefont {Wang}}, \ and\ \bibinfo {author} {\bibfnamefont {G.-L.}\
      \bibnamefont {Long}},\ }\href@noop {} {\bibfield  {journal} {\bibinfo
      {journal} {New Journal of Physics}\ }\textbf {\bibinfo {volume} {22}},\
      \bibinfo {pages} {093009} (\bibinfo {year} {2020})}\BibitemShut {NoStop}%
    \bibitem [{\citenamefont {Chen}\ and\ \citenamefont
      {Clerk}(2014)}]{chen2014photon}%
      \BibitemOpen
      \bibfield  {author} {\bibinfo {author} {\bibfnamefont {W.}~\bibnamefont
      {Chen}}\ and\ \bibinfo {author} {\bibfnamefont {A.~A.}\ \bibnamefont
      {Clerk}},\ }\href@noop {} {\bibfield  {journal} {\bibinfo  {journal}
      {Physical Review A}\ }\textbf {\bibinfo {volume} {89}},\ \bibinfo {pages}
      {033854} (\bibinfo {year} {2014})}\BibitemShut {NoStop}%
    \bibitem [{\citenamefont {Jing}\ \emph {et~al.}(2015)\citenamefont {Jing},
      \citenamefont {{\"O}zdemir}, \citenamefont {Geng}, \citenamefont {Zhang},
      \citenamefont {L{\"u}}, \citenamefont {Peng}, \citenamefont {Yang},\ and\
      \citenamefont {Nori}}]{jing2015optomechanically}%
      \BibitemOpen
      \bibfield  {author} {\bibinfo {author} {\bibfnamefont {H.}~\bibnamefont
      {Jing}}, \bibinfo {author} {\bibfnamefont {{\c{S}}.~K.}\ \bibnamefont
      {{\"O}zdemir}}, \bibinfo {author} {\bibfnamefont {Z.}~\bibnamefont {Geng}},
      \bibinfo {author} {\bibfnamefont {J.}~\bibnamefont {Zhang}}, \bibinfo
      {author} {\bibfnamefont {X.-Y.}\ \bibnamefont {L{\"u}}}, \bibinfo {author}
      {\bibfnamefont {B.}~\bibnamefont {Peng}}, \bibinfo {author} {\bibfnamefont
      {L.}~\bibnamefont {Yang}}, \ and\ \bibinfo {author} {\bibfnamefont
      {F.}~\bibnamefont {Nori}},\ }\href@noop {} {\bibfield  {journal} {\bibinfo
      {journal} {Scientific reports}\ }\textbf {\bibinfo {volume} {5}},\ \bibinfo
      {pages} {1} (\bibinfo {year} {2015})}\BibitemShut {NoStop}%
    \bibitem [{\citenamefont {Zhang}\ \emph {et~al.}(2012)\citenamefont {Zhang},
      \citenamefont {Li}, \citenamefont {Feng},\ and\ \citenamefont
      {Xu}}]{zhang2012precision}%
      \BibitemOpen
      \bibfield  {author} {\bibinfo {author} {\bibfnamefont {J.-Q.}\ \bibnamefont
      {Zhang}}, \bibinfo {author} {\bibfnamefont {Y.}~\bibnamefont {Li}}, \bibinfo
      {author} {\bibfnamefont {M.}~\bibnamefont {Feng}}, \ and\ \bibinfo {author}
      {\bibfnamefont {Y.}~\bibnamefont {Xu}},\ }\href@noop {} {\bibfield  {journal}
      {\bibinfo  {journal} {Physical Review A}\ }\textbf {\bibinfo {volume} {86}},\
      \bibinfo {pages} {053806} (\bibinfo {year} {2012})}\BibitemShut {NoStop}%
    \bibitem [{\citenamefont {Guo}\ \emph {et~al.}(2014)\citenamefont {Guo},
      \citenamefont {Li}, \citenamefont {Nie},\ and\ \citenamefont
      {Li}}]{guo2014electromagnetically}%
      \BibitemOpen
      \bibfield  {author} {\bibinfo {author} {\bibfnamefont {Y.}~\bibnamefont
      {Guo}}, \bibinfo {author} {\bibfnamefont {K.}~\bibnamefont {Li}}, \bibinfo
      {author} {\bibfnamefont {W.}~\bibnamefont {Nie}}, \ and\ \bibinfo {author}
      {\bibfnamefont {Y.}~\bibnamefont {Li}},\ }\href@noop {} {\bibfield  {journal}
      {\bibinfo  {journal} {Physical Review A}\ }\textbf {\bibinfo {volume} {90}},\
      \bibinfo {pages} {053841} (\bibinfo {year} {2014})}\BibitemShut {NoStop}%
    \bibitem [{\citenamefont {Ojanen}\ and\ \citenamefont
      {B{\o}rkje}(2014)}]{ojanen2014ground}%
      \BibitemOpen
      \bibfield  {author} {\bibinfo {author} {\bibfnamefont {T.}~\bibnamefont
      {Ojanen}}\ and\ \bibinfo {author} {\bibfnamefont {K.}~\bibnamefont
      {B{\o}rkje}},\ }\href@noop {} {\bibfield  {journal} {\bibinfo  {journal}
      {Physical Review A}\ }\textbf {\bibinfo {volume} {90}},\ \bibinfo {pages}
      {013824} (\bibinfo {year} {2014})}\BibitemShut {NoStop}%
    \bibitem [{\citenamefont {Liu}\ \emph {et~al.}(2015)\citenamefont {Liu},
      \citenamefont {Xiao}, \citenamefont {Luan},\ and\ \citenamefont
      {Wong}}]{liu2015optomechanically}%
      \BibitemOpen
      \bibfield  {author} {\bibinfo {author} {\bibfnamefont {Y.-C.}\ \bibnamefont
      {Liu}}, \bibinfo {author} {\bibfnamefont {Y.-F.}\ \bibnamefont {Xiao}},
      \bibinfo {author} {\bibfnamefont {X.}~\bibnamefont {Luan}}, \ and\ \bibinfo
      {author} {\bibfnamefont {C.~W.}\ \bibnamefont {Wong}},\ }\href@noop {}
      {\bibfield  {journal} {\bibinfo  {journal} {Science China Physics, Mechanics
      \& Astronomy}\ }\textbf {\bibinfo {volume} {58}},\ \bibinfo {pages} {1}
      (\bibinfo {year} {2015})}\BibitemShut {NoStop}%
    \bibitem [{\citenamefont {Wanjura}\ \emph {et~al.}(2020)\citenamefont
      {Wanjura}, \citenamefont {Brunelli},\ and\ \citenamefont
      {Nunnenkamp}}]{wanjura2020topological}%
      \BibitemOpen
      \bibfield  {author} {\bibinfo {author} {\bibfnamefont {C.~C.}\ \bibnamefont
      {Wanjura}}, \bibinfo {author} {\bibfnamefont {M.}~\bibnamefont {Brunelli}}, \
      and\ \bibinfo {author} {\bibfnamefont {A.}~\bibnamefont {Nunnenkamp}},\
      }\href@noop {} {\bibfield  {journal} {\bibinfo  {journal} {Nature
      communications}\ }\textbf {\bibinfo {volume} {11}},\ \bibinfo {pages} {1}
      (\bibinfo {year} {2020})}\BibitemShut {NoStop}%
    \bibitem [{\citenamefont {Abdo}\ \emph {et~al.}(2013)\citenamefont {Abdo},
      \citenamefont {Sliwa}, \citenamefont {Frunzio},\ and\ \citenamefont
      {Devoret}}]{abdo2013directional}%
      \BibitemOpen
      \bibfield  {author} {\bibinfo {author} {\bibfnamefont {B.}~\bibnamefont
      {Abdo}}, \bibinfo {author} {\bibfnamefont {K.}~\bibnamefont {Sliwa}},
      \bibinfo {author} {\bibfnamefont {L.}~\bibnamefont {Frunzio}}, \ and\
      \bibinfo {author} {\bibfnamefont {M.}~\bibnamefont {Devoret}},\ }\href@noop
      {} {\bibfield  {journal} {\bibinfo  {journal} {Physical Review X}\ }\textbf
      {\bibinfo {volume} {3}},\ \bibinfo {pages} {031001} (\bibinfo {year}
      {2013})}\BibitemShut {NoStop}%
    \bibitem [{\citenamefont {McDonald}\ \emph {et~al.}(2018)\citenamefont
      {McDonald}, \citenamefont {Pereg-Barnea},\ and\ \citenamefont
      {Clerk}}]{mcdonald2018phase}%
      \BibitemOpen
      \bibfield  {author} {\bibinfo {author} {\bibfnamefont {A.}~\bibnamefont
      {McDonald}}, \bibinfo {author} {\bibfnamefont {T.}~\bibnamefont
      {Pereg-Barnea}}, \ and\ \bibinfo {author} {\bibfnamefont {A.}~\bibnamefont
      {Clerk}},\ }\href@noop {} {\bibfield  {journal} {\bibinfo  {journal}
      {Physical Review X}\ }\textbf {\bibinfo {volume} {8}},\ \bibinfo {pages}
      {041031} (\bibinfo {year} {2018})}\BibitemShut {NoStop}%
    \bibitem [{\citenamefont {Turner}\ and\ \citenamefont
      {Stolen}(1981)}]{turner1981fiber}%
      \BibitemOpen
      \bibfield  {author} {\bibinfo {author} {\bibfnamefont {E.}~\bibnamefont
      {Turner}}\ and\ \bibinfo {author} {\bibfnamefont {R.}~\bibnamefont
      {Stolen}},\ }\href@noop {} {\bibfield  {journal} {\bibinfo  {journal} {Optics
      letters}\ }\textbf {\bibinfo {volume} {6}},\ \bibinfo {pages} {322} (\bibinfo
      {year} {1981})}\BibitemShut {NoStop}%
    \end{thebibliography}
%

\end{document}